\documentclass[preprint,12pt]{elsarticle}

\usepackage{amssymb}
\usepackage{amsthm}
\usepackage[bookmarksnumbered, pdfstartview=FitH,colorlinks,urlcolor=blue, citecolor=blue,linkcolor=blue,] {hyperref}

\usepackage{lineno}
\usepackage{color}
\usepackage{mathtools}

\usepackage{subfigure}
\usepackage{caption}

\journal{Nuclear Instruments and Methods in Physics Research Section A}

\usepackage{setspace}
\usepackage{upgreek}
\usepackage[bookmarksnumbered, pdfstartview=FitH,colorlinks,urlcolor=blue, citecolor=blue,linkcolor=blue,] {hyperref}

\begin{document}
\begin{doublespacing}
\begin{frontmatter}
%%\captionsetup[figure]{labelfont={bf}, labelformat={default}, labelsep=period,name={Fig.}}



%\title{Simulating MCP secondary electron avalanche process by Geant4}
\title{Simulating the secondary electron avalanche of MCP by Geant4}


	\author[a,b]{~Huaxing Peng}
	\author[a,b]{~Baojun Yan\footnote{Email: yanbj@ihep.ac.cn}}
	\author[a,c]{~Han Miao\footnote{Email: miaohan@ihep.ac.cn}}
	\author[a,b,c]{~Shulin Liu}
	\author[a,b]{~Binting Zhang}

	\address[a]{\it Institute of High Energy Physics, Beijing 100049, People's Republic of China}
	\address[b]{\it State Key Laboratory of Particle Detection and Electronics, Beĳing, 100049, People's Republic of China}
	\address[c]{\it University of Chinese Academy of Sciences, Beijing 100049, People's Republic of China}

%%\author{}

%%\affiliation{organization={},%Department and Organization

\begin{abstract}

%The curved channel Microchannel Plates(MCP) represent a significant advancement in inhibiting ion-feedback, which is the major cause of after-pulse in output signal of MCPs. Many high sensitive experiments, such as neutrino detection, require as low noise as possible. The conventional straight channel MCP is inevitable to introduce the after-pulses, which are the major source of noise. Normally, couple two straight MCPs in cascade and combine the channels into a "V" shape can effectively suppress ion-feedback, known as chevron MCPs. However, due to the limitation manufacture technology, the gap always exist. It will worsen the resolution and peak-to-valley ratio. Different with the joint MCPs that is mentioned above, the curved MCPs exhibit superior characteristics. Based on the Geant4 Monte Carlo simulation framework, we investigate how the geometrical parameter of curved channel MCP influence the gain. Consequently, when the MCP thickness and operating voltage are fixed, an optimum pore radius exists to reach maximum gain. Additionally, The simulation of electron tracks in MCP channels reveals that the average acceleration distance before incident inner wall is approximately 20 $\upmu \mathrm{m}$ for a curved MCP with applied voltage 950 V, length-diameter-ratio 80:1, and pore diameter 20 $\upmu \mathrm{m}$

Nowadays, Microchannel Plate (MCP), as a kind of electron multipliers based on the secondary electron emission, is widely used in many high-sensitive experiments, such as neutrino detection, which require the noise to be as low as possible, while the conventional straight-channel MCP will inevitably have ion feedback, resulting in the sequential after-pulses being the major source of noise. Normally, the problem can be effectively avoided by coupling two straight-channel MCPs in cascade and combining the channels into a ``V'' shape known as chevron MCPs, but this method is limited by the manufacturing techniques due to the unavoidable gap between the two pieces that will worsen the resolution and peak-to-valley ratio. However, the ion feedback can be inhibited significantly for MCPs with curved channels. Based on Geant4, we investigate how the geometrical parameters of curved-channel MCP influence the gain and get the optimum pore diameter for an MCP to reach the maximum gain with fixed thickness and applied voltage. Additionally, the track-by-track simulation reveals that the average acceleration distance of a secondary electron inside the curved-channel is approximately 20~$\mu$m when the applied voltage, length-to-diameter ratio and pore diameter are 950~V, 50:1 and 20~$\mu$m, respectively.

\end{abstract}



\begin{keyword}



MCP \sep Geant4 \sep Monte Carlo \sep Furman model \sep Secondary electron emission

\end{keyword}

\end{frontmatter}


\section{Introduction}
\label{sec:introduction}

Microchannel Plate (MCP), as a kind of electron multipliers based on the secondary electron emission, is widely used in multiple fields including high energy and nuclear physics~\cite{you1800application, gao2023pos, orlov2019uv, chang2016r}, mass spectrometry~\cite{nishiguchi2006development, klumpar2001time} and image intensifier~\cite{tremsin2009detection, fraser1993x}, because of the 2-dimensional spatial resolution provided by millions of array channels and much higher time resolution than traditional dynode electron multipliers originated from the much smaller thickness of less than 1~mm. Nowadays, the strict demand for the signal-to-noise ratio of single-electron detection in the newly proposed high-energy physics experiments encourages researchers to find solutions for the noise from the after-pulses~\cite{abusleme2022mass}. The after-pulses are from the molecules ionized by the electrons during the avalanche, which will be accelerated towards the cathode by the applied electric field inside the channel, then hit the cathode, and finally produce additional secondary electrons, leading to additional multiplications and resulting in additional output pulses about a few nanoseconds to tens of microseconds after the initial pulse, depending on the gas molecule, the dimension of the glass bulb and the strength of the applied electric field~\cite{abusleme2022mass}. Usually, the time of the after-pulses of 20-inch MCP-PMTs ranges from 500 to 20,000~ns after an initial pulse, while it will be in the order of 6-10~ns for the traditional, very compact MCP-PMTs. The most direct method to suppress the after-pulses is increasing the vacuum degree of the MCP working environment considering the generation mechanism. However, limited by the present techniques, the vacuum degree cannot be increased infinitely, so that other ingenious methods have to be found. A useful method is combining two pieces of straight-channel MCPs to make the channels shaped like ``V'', so-called chevron MCP~\cite{guo2021numerical, leskovar1977microchannel, medley1981response}, so that the ions are not likely to track back to the cathode. This method is limited by the manufacture craft since a gap will always exist between the two pieces and the secondary electrons will be diffused when they go through the gap, leading to the deterioration of peak-to-valley ratio and resolution. Therefore, the natural idea is to make a single MCP with curved channels. Early papers introduced the development and test results of curved-channel MCP, concluding that the curved-channel MCP is superior in resolution, gain and the suppression of ion feedback~\cite{10.1063/1.1136749, 10.1117/1.JATIS.2.3.030901}. However, because of the complex procedures to prepare a curved-channel MCP, such devices cannot be bought for scientific research from the manufacturers. In this paper, we mainly focus on studying the optimal design of curved-channel MCP using Geant4 simulation.

Experimentation is the most straightforward but usually not an efficient or convenient method for most researchers. On the theoretical aspect, limited by the complexity of secondary electron emission and avalanche amplification, many phenomenological models simplify relevant physics processes so that only a portion of reality~\cite{loubet1973mechanism, eberhardt1979gain} can be described, although the corresponding researches started in a very early time~\cite{baroody1950theory, adams1966mechanism}. With the development of computer performance, Monte Carlo simulation has become an important tool in many scientific fields, which also benefits the study of MCP. Commercial softwares like CST Studio Suite~\cite{CST} and SIMION~\cite{SIMION} have already been used in the simulation of MCP, while an open-source software is still needed and will always be welcome in the relevant research. In this paper, based on the open-source toolkit Geant4\cite{Geant4}, we simulate the entire multiplication of electrons inside the MCP with adding the process of secondary electron emission referring to the minorly modified Furman model~\cite{furman2002probabilistic} to meet our requirement. With incorporating the similar method in this work, we have developed an independent program for the simulation of electron multipliers represented by MCP.~\cite{2023arXiv231005122M}

%In Section~\ref{sec:theory} we introduce the basic theory of secondary electron emission and discuss secondary electrons behave in the low-energy range. At the end of Section~\ref{sec:theory}, we make some  modifications to the Furman model. In Section~\ref{sec:simulation_result}, we compare the experimental data with the simulation result, and study the effects of various parameters on the performance of curved channel MCP. In section~\ref{sec:track_simulation}, we study the electron track in curved channel MCP and analyze the results. And last, we summarize the entire article.

This paper is organized in the following manner: Section~\ref{sec:theory} introduces the phenomenological model of secondary electron emission used in this work and discusses the behavior of the secondary electrons within the low-energy range. Section~\ref{sec:simulation_result} compares the experimental measurements with the simulation result and studies the influence of various parameters on the performance of curved-channel MCP. Section~\ref{sec:track_simulation} is the detailed study of the electron tracks inside the curved-channel MCP. At last, the entire work is summarized in Section~\ref{sec:conclusion}.

\section{Phenomenological model}
\label{sec:theory}
\subsection{Furman model}

%In this paper, we use Furman model \cite{furman2002probabilistic} to simulate secondary electron emission process in MCP. The Furman model is a phenomenological model that subjects to energy conservation. Besides, the model uses probability theory to describe the generation of secondary electrons and is mathematically self-consistent \cite{furman2002probabilistic}. So the Furman model is widely used by Monte Carlo method to simulate Photomultiplier Tube(PMT), MCP detectors\cite{guo2021numerical}. Within the Furman model, the emitted electrons are categorized based on their generation mechanisms, which can be divided into three types, depicted in Figure~\ref{fig:SE_mechanism}.  Backscattered electrons are caused by elastic collision with the atoms of the material. In this process, there is little energy transfer, resulting in backscattered electron emission energy being close to primary electron energy. Rediffused electrons, on the other hand, are generated through inelastic collisions with material atoms. Compared with collisions, inelastic collisions involve significant energy transfer. As a result, the energy of rediffused emission ranges from 0 eV to primary electron energy. True secondary electrons mainly response for electron avalanche in micro-channel. When the primary electrons penetrate material surface, partial energy will deposit in the solid, exciting other electrons. These excited electrons may diffuse to the surface and emit to the free space. 

\begin{figure}[htbp]
	\centering
	\includegraphics[width=0.5\textwidth]{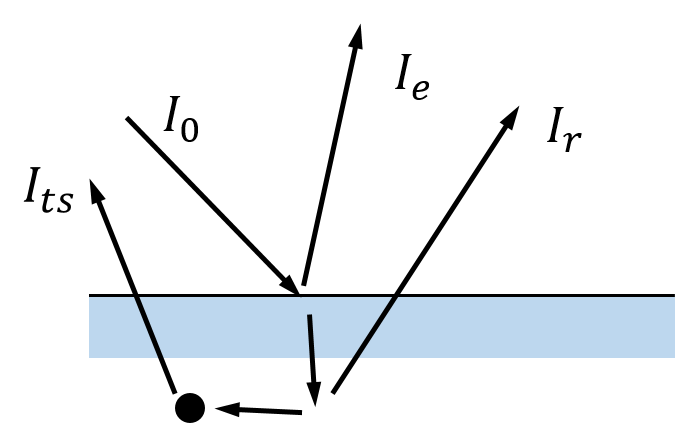}
	\caption{The three types of secondary electron emission processes. $I_0$ represents incident electron, $I_e$ represents the backscattered electron, $I_r$ represents the rediffused electron and $I_{ts}$ represents the true secondary electron.}
	\label{fig:SE_mechanism}
\end{figure}

In this paper, the Furman model~\cite{furman2002probabilistic} is used to simulate the secondary electron emission inside the vacuum electron multipliers like MCP. As a phenomenological model that uses probability theory to describe the production of secondary electrons with keeping energy conservation and being mathematically self-consistent, the Furman model is commonly incorporated in the Monte Carlo simulation of the Photomultiplier Tube (PMT) and MCP~\cite{guo2021numerical}. In the Furman model, the emitted electrons are categorized into three types based on their generation mechanisms, as depicted in Fig.~\ref{fig:SE_mechanism}. Backscattered electrons, of which the energies are close to the incident energy, are from the elastic collisions between the incident electron and the atoms in the material. Rediffused electrons are produced in inelastic scattering. Because of the significant energy transfer, the rediffused electrons energies range is from 0~eV to the incident energy. True secondary electrons, as the main part of the electron avalanche, are from the other electrons excited by the deposited energy of the incident electron, which may be diffused to the surface and finally emit into the free space.

%Secondary electron yield(SEY) and energy spectra of secondary electrons are two important physical quantity that can describe the process of secondary electron emission\cite{bruining2016physics,redhead1968physical, van2022secondary}. The SEY, denoted as $\delta$, quantifies the capacity of a material to emit secondary electrons, defined in Equation~\ref{equ:sey_define}.

The secondary electron emission is modeled by two typical variables: the secondary electron yield (SEY) and the energy spectrum of the secondary electrons~\cite{bruining2016physics,redhead1968physical, van2022secondary}. The SEY, denoted as $\delta$, quantifies the capacity of a material to emit secondary electrons, defined as
\begin{equation}
	\label{equ:sey_define}
	\delta = \frac{I_s}{I_0}
\end{equation}
where $I_0$ and $I_s$ represent the number of incident electrons and the number of emitted electrons from the material surface, respectively. For the electron multipliers, a SEY value larger than 1 is typically required. Additionally, the energy spectra of the emitted secondary electrons is defined as ${\rm d}\delta / {\rm d} E$, where $E$ is the energy of the emitted electrons~\cite{furman2002probabilistic}.

%$I_0$ represents the number of incident electron and $I_s$ represents the number of the emission electron from material surface. . For the electron multiplier device, a SEY value larger than 1 is typically desired. Base on the $\delta$, the emitted-energy spectrum is ${\rm d}\delta / {\rm d} E$\cite{furman2002probabilistic}.

%In the Furman model, we use some empirical formulas to describe SEY and energy spectra of secondary electrons. In fact, these formulas are non-unique\cite{wu2008monte} and the parameters they involve do not have physical significance. However, they are convenient for sampling in Monte Carlo methods and have good fitting to experiment data.

In the Furman model, some empirical formulas are used to describe the SEY and the energy spectrum of secondary electrons. In fact, the formulas are non-unique~\cite{wu2008monte} and usually the parameters involved don't have clear physics meanings, while they are convenient for the sampling in Monte Carlo methods and fit well to the experimental measurements.

%Equation~\ref{equ:true_sey},~\ref{equ:elastic_sey} and~\ref{equ:rediff_sey} correspond to true secondary electron yield, backscatter secondary electron yield and rediffused secondary electron yield , respectively\cite{bruining2016physics,redhead1968physical}. In equation~\ref{equ:true_sey}, the $\delta_{t s}\left(E_0, \theta_0\right)$ denotes true secondary electron yield when incident electron energy is $E_0$ and incident angle is $\theta_0$. In the $D(x)$ function, $s$ is a parameter that control curve shape. Furthermore, when $x=1$, the $D(x)$ reaches its maximum value. In other words, when incident electron energy is $\hat{E}\left(\theta_0\right)$, SEY value is maximum $\hat{\delta}\left(\theta_0\right)$, like in Figure~\ref{fig:SEY} .

\begin{figure}[htbp]
	\centering
	\subfigure[Dependence of SEY on incident energy of three types of secondary electrons]{
		\includegraphics[width=0.4\textwidth]{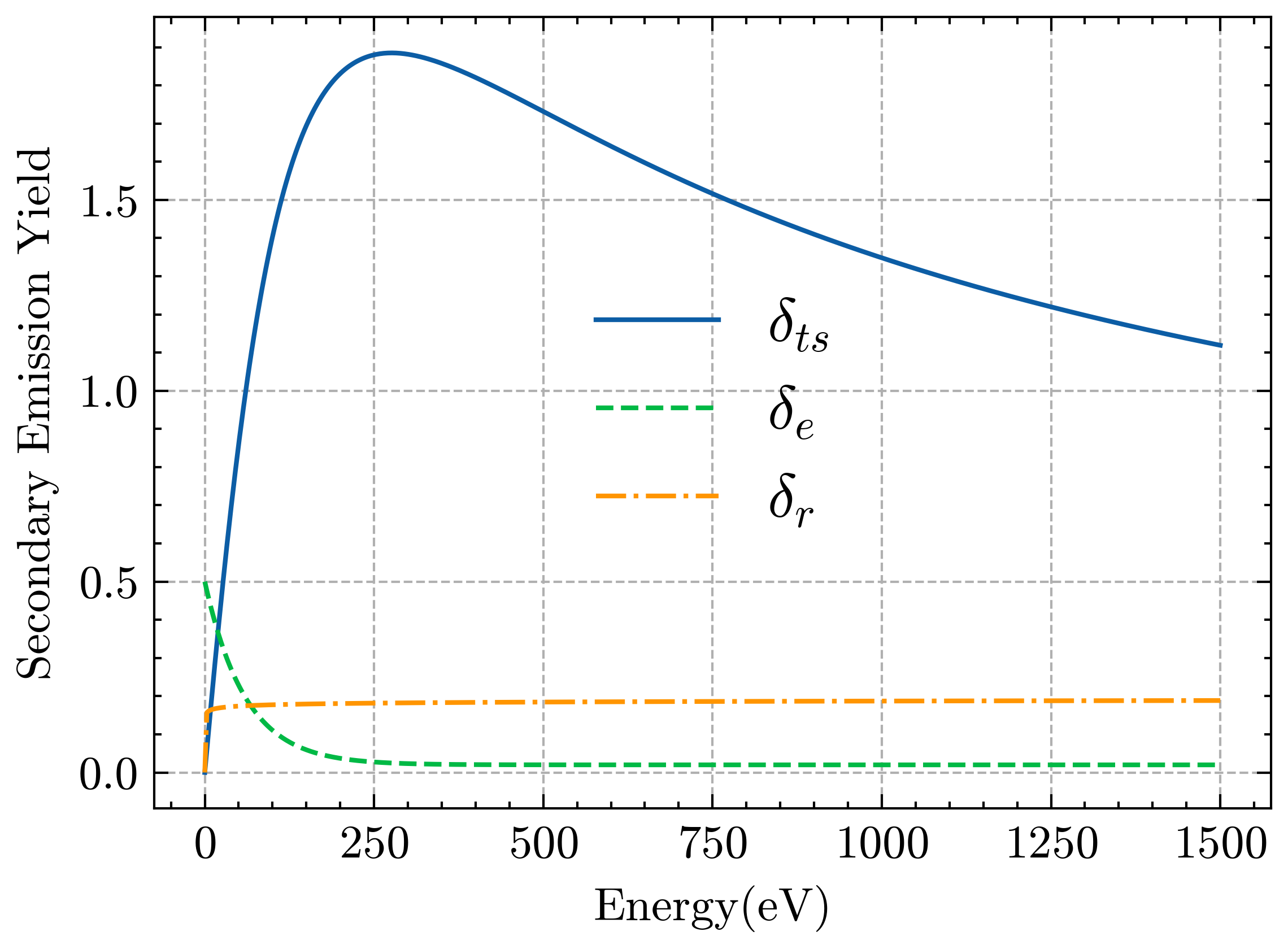} 
		\label{subfig:sey}
	} 
	\subfigure[Dependence of SEY on incident energy of three types of secondary energy within 0 eV to 100 eV]{
		\includegraphics[width=0.4\textwidth]{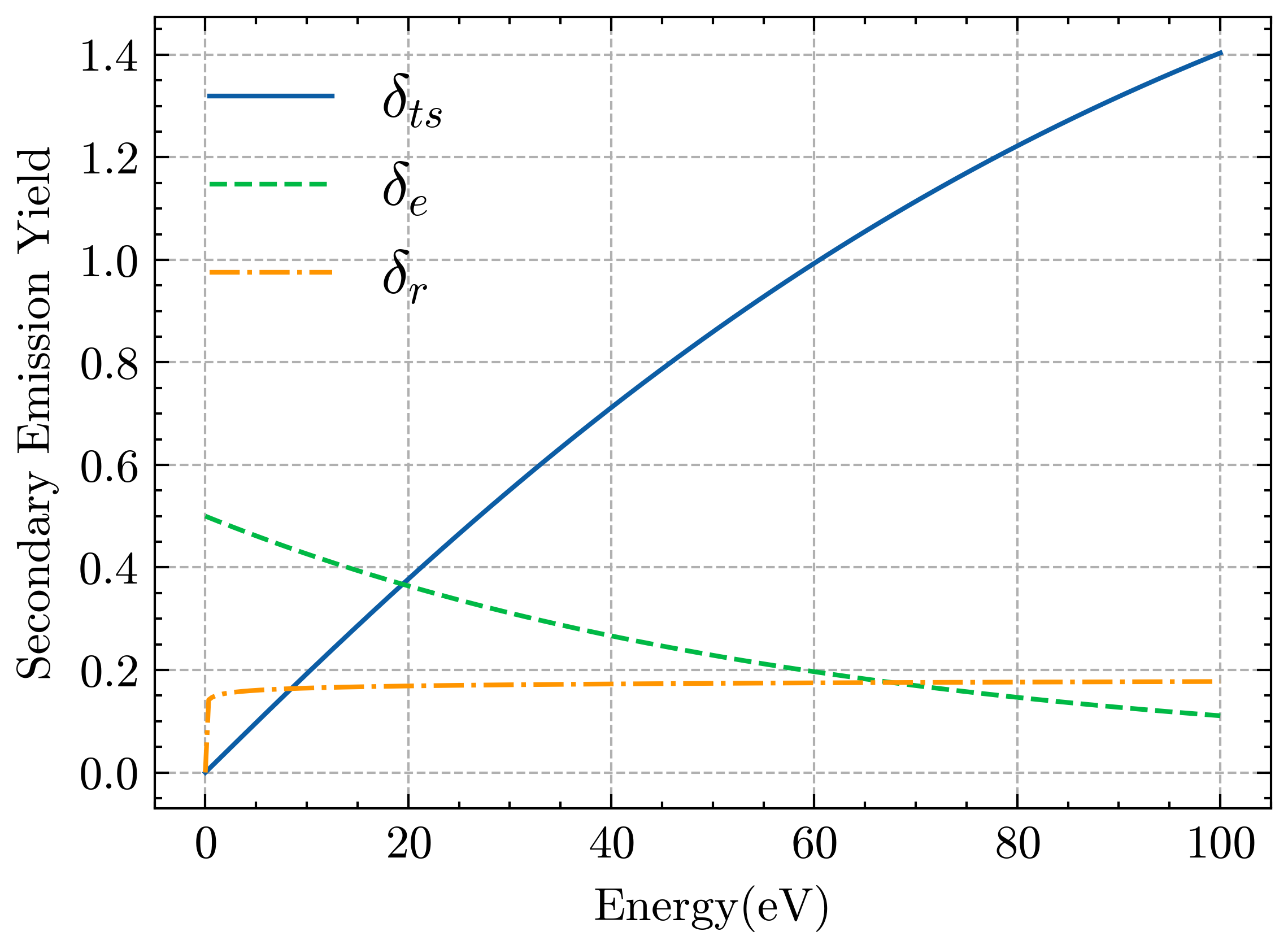}
		\label{subfig:sey_zoom}
	}
	\caption{SEY for different processes described by empirical formulas.}
	\label{fig:SEY}
\end{figure}

\begin{figure}[htbp]
	\centering
	\includegraphics[width=0.5\textwidth]{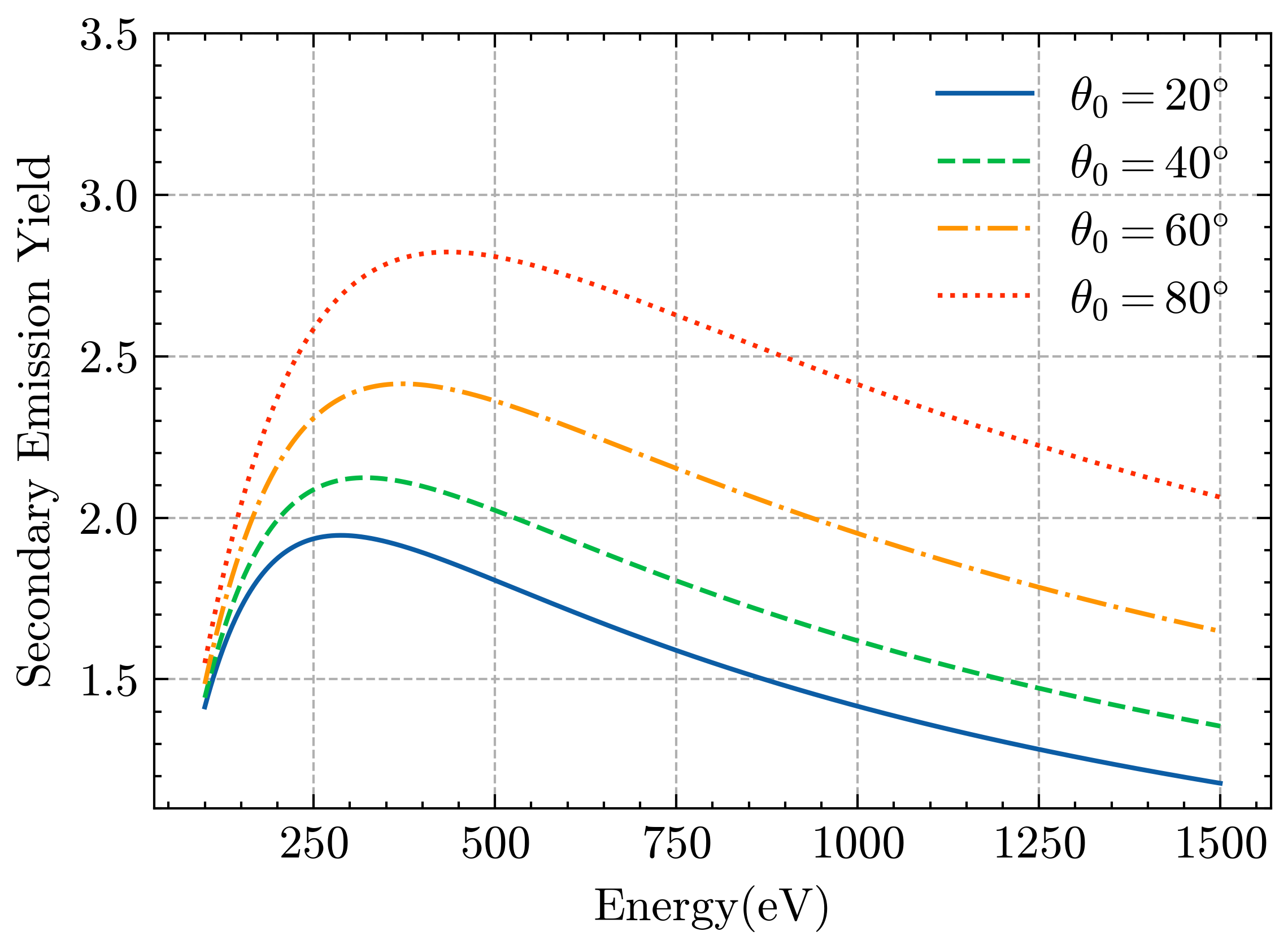}
	\caption{SEY of the true secondary electrons with respect to the incident angles.}
	\label{fig:SEY_angle}
\end{figure}

The SEY of the true secondary electrons can be described as \cite{baroody1950theory, furman1997proceedings, kirby2001secondary}
\begin{equation}
	\label{equ:true_sey}
	\begin{aligned}
		\delta_{t s}\left(E_0, \theta_0\right)&=\hat{\delta}\left(\theta_0\right) D\left[E_0 / \hat{E}\left(\theta_0\right)\right], \\
		D(x)&=\frac{s x}{s-1+x^s},\\
		\hat{\delta}\left(\theta_{0}\right)&=\hat{\delta}_{t s}\left[1+t_{1}\left(1-\cos ^{t_{2}} \theta_{0}\right)\right], \\
		\hat{E}\left(\theta_{0}\right)&=\hat{E}_{t s}\left[1+t_{3}\left(1-\cos ^{t_{4}} \theta_{0}\right)\right],
	\end{aligned}
\end{equation}
where $\delta_{ts}\left(E_0, \theta_0\right)$ is the SEY of the true secondary electrons when the incident energy and angle are $E_0$ and $\theta_0$. The symbol $s$ is a parameter controlling the shape of $D(x)$ that will reach its maximum $\hat{\delta}\left(\theta_0\right)$ when $x=1$, or $E_0 = \hat{E}\left(\theta_0\right)$, as shown in Fig.~\ref{fig:SEY}. $\hat{E}\left(\theta_0\right)$ and $\hat{\delta}\left(\theta_0\right)$ will also be influenced by the incident angle as shown in Eq.~(\ref{equ:true_sey}), where $t_1$, $t_2$, $t_3$ and $t_4$ are all obtained from the fit to the experimental measurements. Fig.~\ref{fig:SEY_angle} shows the SEY with respect to the incident angle.

%Under the specify incident energy, SEY also will be changed by the incident angle. It can be described by equation~\ref{equ:angle_sey}. $t_1$, $t_2$, $t_3$ and $t_4$ are all fit parameters. The Figure~\ref{fig:SEY_angle} shows SEY curves in different incident angle.
%\begin{equation}
%	\label{equ:angle_sey}
%	\begin{aligned}
%		&\hat{\delta}\left(\theta_{0}\right)=\hat{\delta}_{t s}\left[1+t_{1}\left(1-\cos ^{t_{2}} \theta_{0}\right)\right] \\
%		&\hat{E}\left(\theta_{0}\right)=\hat{E}_{t s}\left[1+t_{3}\left(1-\cos ^{t_{4}} \theta_{0}\right)\right]
%	\end{aligned}
%\end{equation}

%In equation~\ref{equ:elastic_sey}, $\delta_e$ denotes backscatter electron yield, while in equation~\ref{equ:rediff_sey}, $\delta_r$ represents rediffused electron yield. All parameters in these equations, except for $E_0$, which is the incident energy, are fitting parameters. From Figure~\ref{subfig:sey} and Figure~\ref{subfig:sey_zoom}, we can find that the proportion of backscatter electron is high in low energy region and decrease to $P_{1, e}(\infty)$ as the incident energy increase. The rediffused electron yield remains relatively flat in the region of interest.
%\begin{equation}
%	\label{equ:elastic_sey}
%	\delta_e\left(E_0, 0\right)=P_{1, e}(\infty)+\left[\hat{P}_{1, e}-P_{1, e}(\infty)\right] e^{-\left(\left|E_0-\hat{E}_e\right| / W\right)^p / p}
%\end{equation}

The SEYs of the backscattered and rediffused electrons are described by
\begin{equation}
        \label{equ:elastic_sey}
        \delta_e\left(E_0, 0\right)=P_{1, e}(\infty)+\left[\hat{P}_{1, e}-P_{1, e}(\infty)\right] e^{-\left(\left|E_0-\hat{E}_e\right| / W\right)^p / p},
\end{equation}
and
\begin{equation}
        \label{equ:rediff_sey}
        \delta_r\left(E_0, 0\right)=P_{1, r}(\infty)\left[1-e^{-\left(E_0 / E_r\right)^r}\right],
\end{equation}
where all the parameters, except the $E_0$, are obtained from the fit of experimental measurements. With the incident energy increasing, the SEYs of the backscattered and rediffused electrons will be close to $P_{1, e}(\infty)$ and $P_{1, r}(\infty)$. As shown in Fig.~\ref{fig:SEY}, the SEY of backscattered electrons is high within the low energy region and decreases slowly with the incident energy increasing, while the SEY curve of the rediffused electrons is relatively flat.

%\begin{equation}
%	\label{equ:rediff_sey}
%	\delta_r\left(E_0, 0\right)=P_{1, r}(\infty)\left[1-e^{-\left(E_0 / E_r\right)^r}\right]
%\end{equation}

\subsection{Electric field calculation}

For the simulation of MCP, the accuracy of the electric field inside the channels has a significant effect on the results. In principle, the electric field can be solved by the generalized Poisson equation with fixed boundary and initial conditions as
\begin{equation}
	\label{equ:generial_poisson}
	\nabla [\varepsilon(\mathbf{r}) \nabla V(\mathbf{r})]=-\frac{\rho(\mathbf{r})}{\varepsilon_0}
\end{equation}
where $V$ is the potential we aim to solve and $\rho$ is the charge density that is 0 in the case of MCP since no free charge distributes in the domain of interest. The symbol $\varepsilon_0$ is the vacuum permittivity and $\varepsilon(\mathbf{r})$ denotes the relative permittivity which could be a function of the position if the material is nonuniform. The most common numerical method to solve this equation is the Finite Difference Method (FDM) with structured grids, while this method may not work well in the case when the dielectric material suddenly changes so that the permittivity $\varepsilon$ becomes discontinuous. Therefore, the Finite Element Method (FEM) is chosen in this work, which converts the generalized Poisson equation into a weak form as
\begin{equation}
        \begin{aligned}
                \label{equ:3d_weak_formula}
                \int_\Omega(\nabla v)^{\mathrm{T}} \varepsilon \nabla V \mathrm{~d} \Omega & =-\int_{S_k} v E_n \mathrm{~d} S-\int_{S_g} v \nabla V \mathrm{~d} S+\int_V v \rho \mathrm{~d} \Omega, \\
                V & =\phi \text { on } S_g,
        \end{aligned}
\end{equation}
where $\Omega$ represents the domain to be solved, $S_k$ and $S_g$ are the boundaries of the domain, and $E_n$ denotes the component of the electric field along the normal direction of the plane $S$. The symbol $v$ is called basis functions in FEM~\cite{mazumder2015numerical}. Commonly, FEM requires subdividing the domain of interest into an unstructured grid, which can adapt to complex geometric domains. An open-source software Gmsh~\cite{geuzaine2009gmsh} is incorporated in this work, containing many C++ and Python interfaces that we can easily embed into our code.

%\begin{equation}
%	\begin{aligned}
%		\label{equ:3d_weak_formula}
%		\int_\Omega(\nabla v)^{\mathrm{T}} \epsilon \nabla V \mathrm{~d} \Omega & =-\int_{S_k} v E_n \mathrm{~d} S-\int_{S_g} v \nabla V \mathrm{~d} S+\int_V v \rho \mathrm{~d} \Omega \\
%		V & =\phi \text { on } S_g
%	\end{aligned}
%\end{equation}

\subsection{Treatment in low energy region}

%In fact, regardless of the true secondary electron emission mechanism, according to conservation of energy, there must be a certain energy threshold for the excited electron to overcome the surface potential energy of the material. This can be compared to the photoelectric effect. As we know that when photon frequency below threshold frequency, there is no electron can overcome surface potential to emission. Similarly, the true secondary electron can only be emitted when the primary electron energy is above a specify value.

It is imaginable that there should be a threshold of the incident energy for the produced secondary electrons inside the material to escape into the free space, similar to the energy threshold in the photoelectric effect, so that the true secondary electrons can only be emitted when the incident energy is above a certain value.

%In Furman model, equation~\ref{equ:true_sey} shows good agreement with experimental data for incident energy above 100 eV. However, based on the above understanding, we can conclude that the true secondary electron curve does not pass through the origin (0, 0) point. As the incident energy decrease, the true secondary electron yield decreases and reaches 0 when the incident energy decrease below a certain value. Therefore, we need to modify the SEY curve. 

Eq.~(\ref{equ:true_sey}) in the Furman model shows great agreement with experimental measurements for incident energy above 100~eV, while the original curve passes through $(0,0)$ so that it has to be modified to meet the requirement of energy conservation as mentioned above.

%From Figure~\ref{fig:SEY}, we can find that in the high energy region (incident energy > 100eV ), backscattered electron yield and rediffused electron yield are approximately constant. This suggests that the shape of true secondary electron yield curve is similar to the total secondary electron yield curve above 100 eV. We can use equation~\ref{equ:true_sey} to fit the total yield and obtain $\delta_{total}$. To acquire true yield, we subtract $\delta_e$ from $\delta_{total}$ then subtract $\delta_r$. Finally, we get the new equation~\ref{equ:new_true_sey} to describe true secondary yield. The function curve is showed in Figure~\ref{fig:modified_SEY}.

\begin{figure}[htbp]
	\centering
	\subfigure[Modified SEY curve of the three types of secondary electrons.]{
		\includegraphics[width=0.4\textwidth]{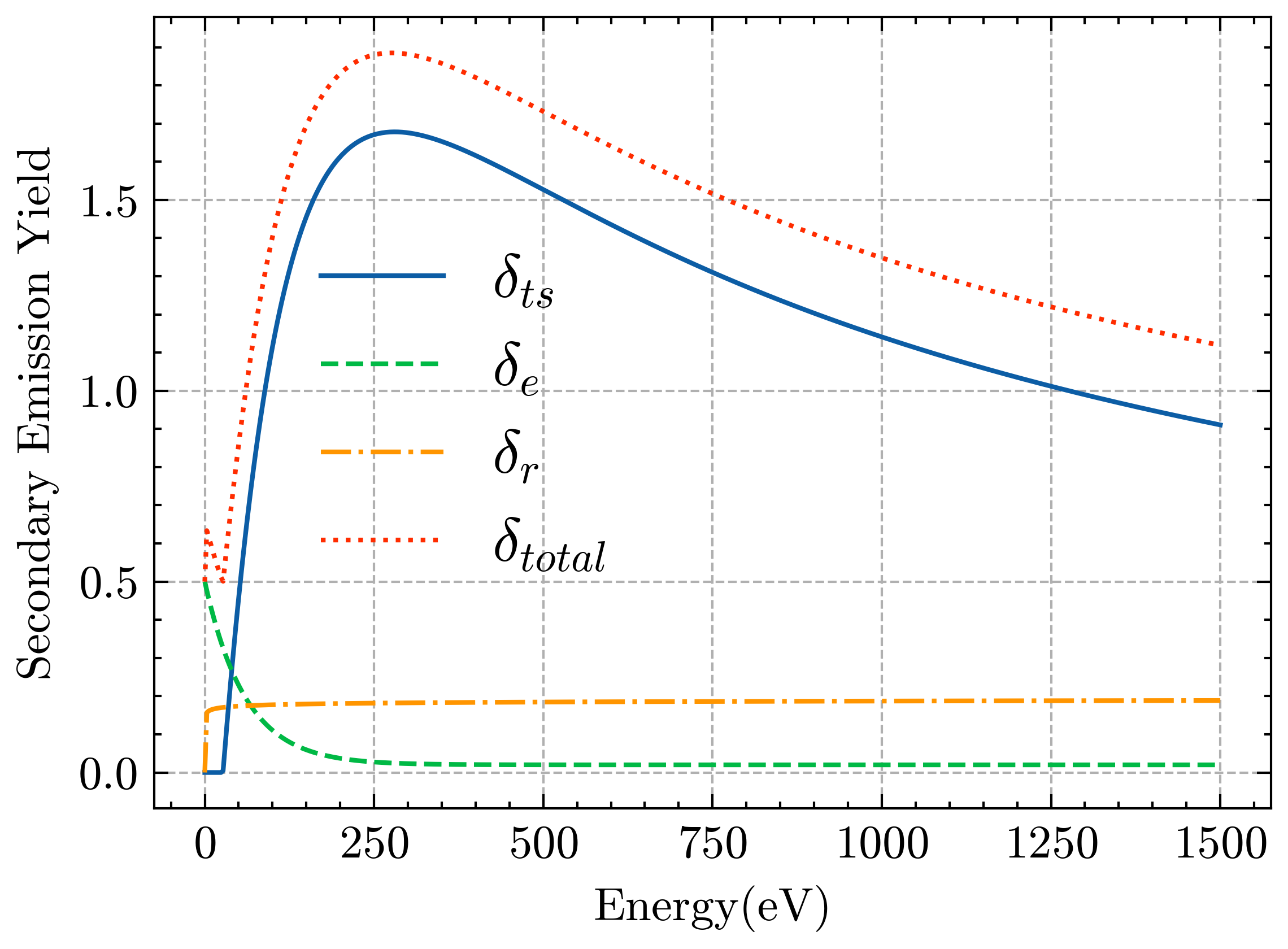} 
		\label{subfig:modify_sey}
	} 
	\subfigure[Modified SEY curve of the three types of secondary electrons with incident energy within 0~eV to 100~eV.]{
		\includegraphics[width=0.4\textwidth]{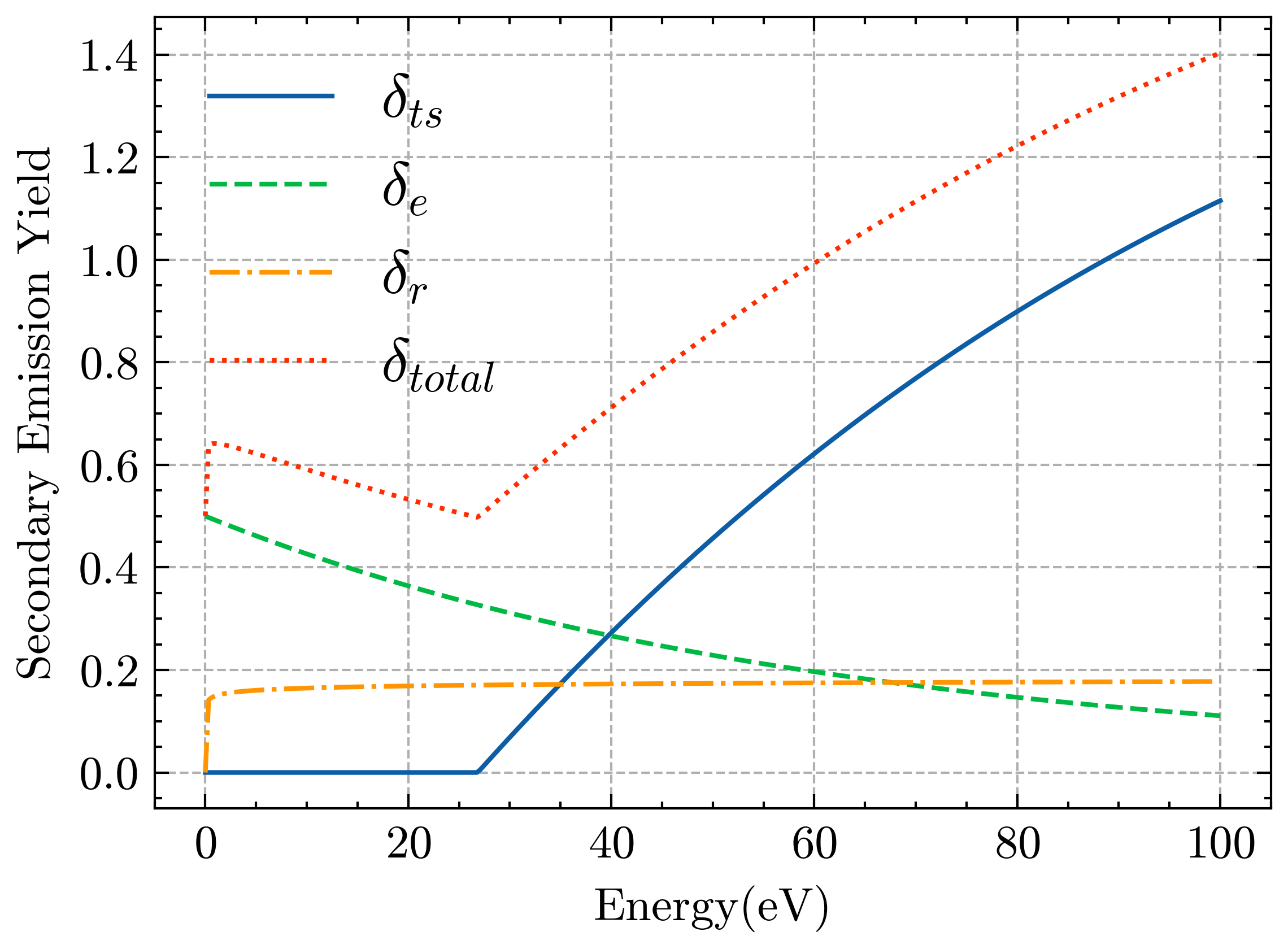}
		\label{subfig:modify_sey_zoom}
	}
	\caption{Modified SEY curves in the Furman model.}
	\label{fig:modified_SEY}
\end{figure}

According to Fig.~\ref{fig:SEY}, the SEYs of the backscattered and rediffused electrons are approximately constant in comparison with the true secondary electrons when the incident energy larger than 100~eV, which suggests that the total SEY curve ($\delta_{total}$) can be fitted using the SEY function of the true secondary electrons as Eq.~(\ref{equ:true_sey}) with additionally adding a constant item. Then, as shown in Eq.~(\ref{equ:new_true_sey}), the SEY of the true secondary electrons is obtained again, denoted as $\delta^{\prime}_{ts}$, by subtracting the SEYs of the backscattered ($\delta_{e}$) and rediffused electrons ($\delta_{r}$) calculated by Eq.~(\ref{equ:elastic_sey}) and (\ref{equ:rediff_sey}) from $\delta_{total}$ when the original SEY of the true secondary electrons $\delta_{ts}$ is larger than 0. And, $\delta^{\prime}_{ts}$ will be 0 if $\delta_{ts}<0$. The modified SEY curves are shown in Fig.~\ref{fig:modified_SEY}.

\begin{equation}
	\label{equ:new_true_sey}
	\delta^{\prime}_{ts} = \begin{cases}
	\delta_{total} - \delta_e - \delta_r  & \quad \delta_{ts} > 0 \\
	0  & \quad \delta_{ts} < 0 \\
\end{cases}
\end{equation}

%In fact, Baglin et al(2001)\cite{baglin2001summary} have measured the copper sample SEY. The most useful experimental data for our study is about the SEY of copper sample ranges from 4 eV to 30 eV. The data show that the SEY value at 4 eV is about 0.7 and the SEY curve tends to a constant or a bit decrease when the incident electron energy changes from 4 eV to 10 eV. Considering the Figure~\ref{fig:modified_SEY}, the SEY curve is slightly decrease under 30 eV, which tendency is similar to the study of Baglin et al(2001). It indirectly support our assumption about low energy region.

The indirect experimental support for our treatment is from the measurements undertaken by Baglin and others~\cite{baglin2001summary}, where the SEY of the copper sample tends to be constant or decrease a little when the incident energy ranges from 4~eV to 10~eV, similar to the slightly decreasing tendency in Fig.~\ref{fig:modified_SEY} obtained under our low-energy treatment.

\section{Simulation of MCP}
\label{sec:simulation_result}
\subsection{Measurement of the SEY and MCP gain}

%To make the simulation results closer to reality, we conducted measurements of the SEY curve for lead glass, which is the main material used in MCP. In our laboratory, we have a spherical vacuum device\cite{wen2018spherical} that can accurately measure SEY of various material, which incident electrons energy ranges from 100 eV to 1500 eV. In Figure~\ref{fig:fit_SEY}, it can be observed that the data points of measured SEY curve are not as smooth as expected. The main reason for this phenomenon is the poor conductivity of the lead glass sample. During the test, it's easy to cause charge accumulation that influences the secondary electron emission. Therefore, compared to good conductors, it's more challenging to accurately measure the SEY of poor conductors. This also bring to another issue. Due to the significant errors encountered in measuring the SEY of lead glass, measuring the secondary electrons energy spectra becomes even more difficult. The energy spectra measurements require the detection of smaller signals, which introduces larger errors. Additionally, in order to acquire different incident energies SEES, we have to measure more data points than SEY measurements. As a result, acquiring the secondary electrons energy spectra of lead glass material becomes a challenging task.

\begin{figure}[htbp]
	\centering
	\includegraphics[width=0.5\textwidth]{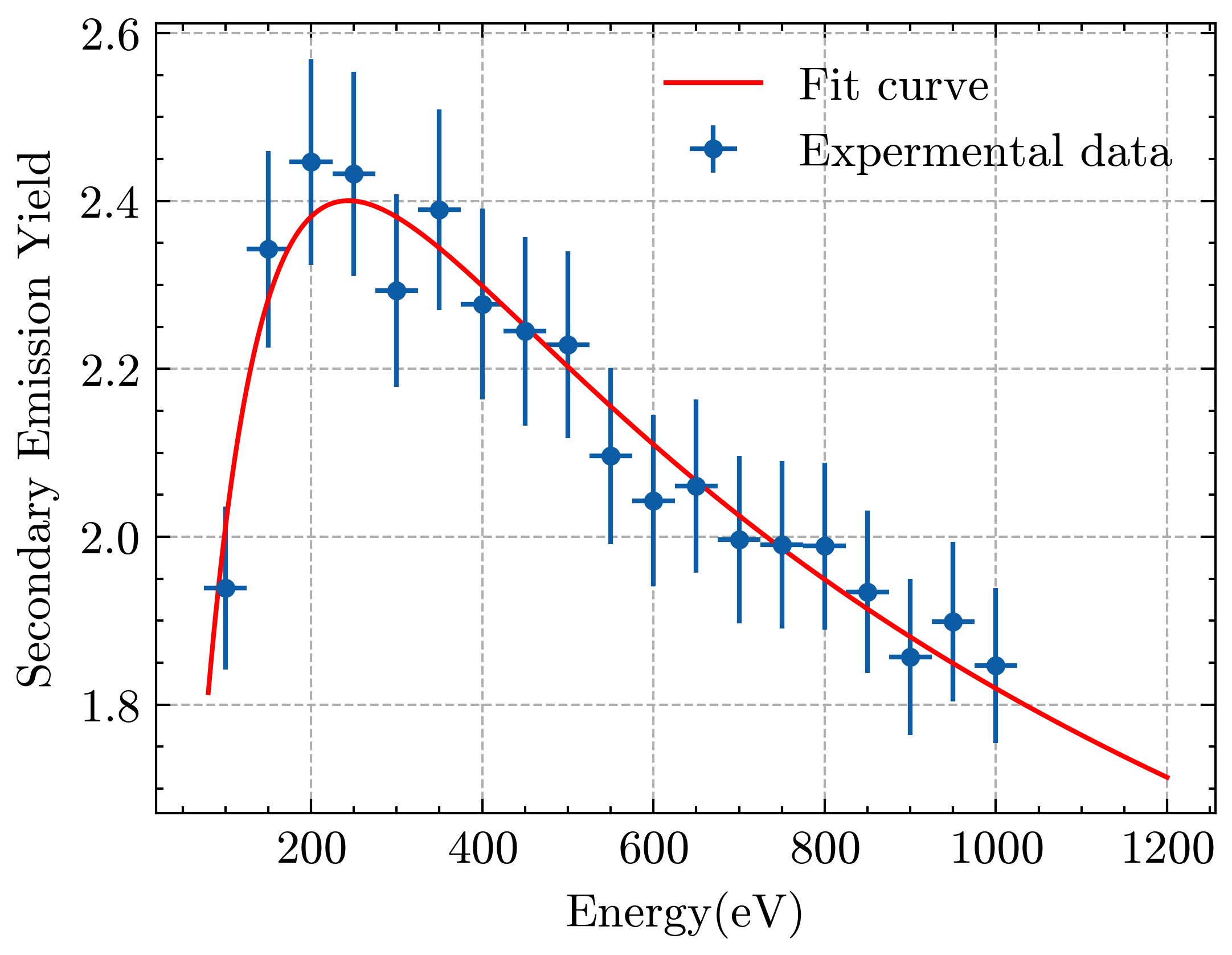}
	\caption{The measured SEY of lead glass overlaid with the best fit using Eq.~(\ref{equ:true_sey}).}
	\label{fig:fit_SEY}
\end{figure}

To make the simulation closer to reality, the measurement of the SEY curve of lead glass, the main material used in MCP, is conducted out using a spherical vacuum device~\cite{wen2018spherical}, which can accurately measure the SEY of various materials with incident energy ranging from 100~eV to 1500~eV. It can be seen that the measured SEY curve in Fig.~\ref{fig:fit_SEY} is not so smooth and has large statistical errors, since the serious charge accumulation of the lead glass, which has poor conductivity in comparison with good conductors, makes the measurement a much more complex mission. The encountered significant statistical errors also make it a really challenging task to measure the energy spectra of the secondary electrons with different incident energies due to the much less statistics we can obtain.

The Eq.~(\ref{equ:true_sey}) is used to fit the experimental measurements shown in Fig.~\ref{fig:fit_SEY} and the corresponding parameters are obtained to be
\begin{equation}
	\label{equ:fit_true_para}
	\begin{aligned}
		s &= 1.39, \\
		\hat{\delta}(0) &= 2.40, \\
		\hat{E}(0) &= 243.71.
	\end{aligned}
\end{equation}

%We have measured the MCP gain at various applied voltage.  The experimental setup is illustrated in Figure~\ref{fig:gain_sketch}. In a vacuum device, a Tantalum filament, the MCP and an anode were placed sequentially. When the system is working, a current $I_{s}$ flows through tantalum filament. Some of electrons emitted from the filament will enter the input pore of MCP. The galvanometer $G_{I}$ measures the input current $I_{in}$. $U_{0}$ applies on MCP. After the avalanche amplification process in MCP channel, we can measure output current from anode. By applying the formula in Equation~\ref{equ:gain_formula}, the MCP gain at the operating voltage was calculated.

\begin{figure}[htbp]
	\centering
	\includegraphics[width=0.8\textwidth]{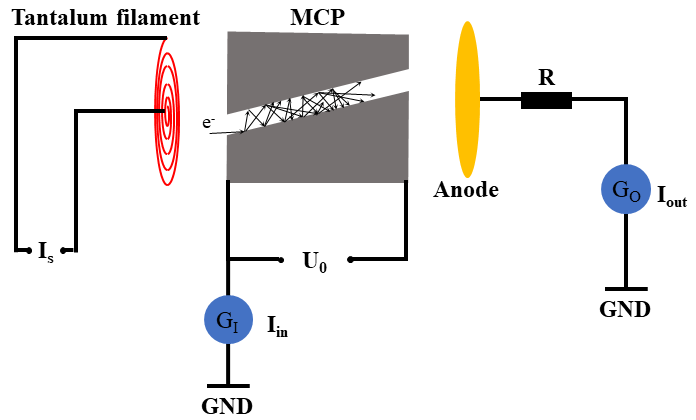}
	\caption{The schematic diagram of the device used to measure the MCP gain.}
	\label{fig:gain_sketch}
\end{figure}

Using the device shown in Fig.~\ref{fig:gain_sketch}, the MCP gains at various applied voltages are measured for the comparison with simulation results. As illustrated in Fig.~\ref{fig:gain_sketch}, a Tantalum filament, the MCP to be measured and an anode are placed sequentially inside a vacuum device. A voltage of $U_0$ is applied to the MCP. When the system is working, a current $I_{s}$ flows through the tantalum filament and some of the electrons emitted from the filament will enter the input pore of the MCP, of which the input current $I_{in}$ will be measured by the galvanometer $G_{I}$. After the avalanche amplification in the MCP channel, the output current $I_{out}$ is measured by the anode and the MCP gain at the applied voltage is then calculated by Eq.~(\ref{equ:gain_formula}).

\begin{equation}
	\label{equ:gain_formula}
	G = \frac{I_{out}}{I_{in}}
\end{equation}

%As mentioned above, it is challenging to directly measure the secondary electron emission spectrum of poor conductor lead glass. However, there are a lot of SEEC data available for other material. The MCP gain characteristic is primarily determined by the material secondary electron mission properties. Therefore, based on the secondary electron energy spectra data of other material and the MCP gain data, by adjusting parameters of energy spectra to iterate simulation process, we can eventually obtain simulation results that are consistent with experimental results. At this point, the energy spectra parameters used in the simulation are close to the real values.

As mentioned above, it is really challenging to directly measure the energy spectra of the secondary electrons of the lead glass, while there are quantities of measured spectra for other materials. Therefore, the initial energy spectrum is obtained by adjusting those of other materials and then used in the original simulation, which will be used to adjust the energy spectrum by comparing the simulation results with experimental measurements. After multiple iterations, the simulated gain curve will converge towards the experimental measurements. Consequently, the energy spectrum will closely approximate the actual case and finally be utilized in our practical simulation.

\subsection{The simulation of MCP gain versus applied voltage}
\label{sec:sim_MCP_gain}
%The gain characteristic of an MCP is not only influenced by the constituent materials but also by its geometric dimensions and applied voltage. Under an optical microscope, we can measure the pore diameter, length-to-diameter ratio of the MCP. By breaking MCP, we can also measure the channel bias angle or other inner geometric dimensions. Based on these actual measurements, in Geant4, the MCP model is established. For the \# 1 MCP, the dimensions are provided in Table~\ref{tab:mcp_1_par}

In this work, two pieces of MCPs are simulated.

The gain of MCP is not only influenced by the constituent materials but also by the geometric dimensions and applied voltage. The pore diameter and length-to-diameter ratio can be measured under an optical microscope, and the tilt angle or other inner geometric dimensions can be measured after breaking the MCP. The geometries of \#1 and \#2 MCP are established in Geant4 based on the actual measurements and provided in Table~\ref{tab:mcp_1_par}.

\begin{table}[htbp]
	\centering
	\doublespacing
	\caption{Geometric parameters of the two pieces of MCPs simulated in this work.}
	\begin{tabular}{ccc}
		\hline
		\hline
		Parameter/Unit   &   \# 1 MCP  & \# 2 MCP\\
		\hline
		\texttt{MCP diameter/mm}    &   25 &24.8\\
		\texttt{MCP thickness/mm} &   0.42 &0.48   \\
		\texttt{Pore diameter/$\upmu$m}    &   10 &6 \\
		\texttt{Tilt angle/$^\circ$}    &   12  &5.5  \\
		\texttt{Body resistance/M$\Omega$}   &   84  &90\\
		\texttt{Input current/pA}  &   107 &162  \\
		\hline
		\hline
	\end{tabular}
	\label{tab:mcp_1_par}
\end{table}

%After above preliminary steps, we can proceed with running the simulation framework and record calculation results in a root file. In Geant4, we place a sensitive detector(SD) adjacent to the output face of the MCP to capture the output electron kinetic data. When a particle enter the SD, we terminate the particle and halt any further calculations related to it in order to conserve computing resources. So no particle can go through the SD, and we regard all particles are absorbed by SD completely.

In the simulation, the kinematic information of all the particles that are able to reach the exit of the channel is recorded without considering the detection efficiency, fluctuation, or other relevant issues of the readout system, so the MCP gain will simply be calculated by dividing the number of the electrons at the exit by the number of the electrons shot towards the MCP.

%We simulate \# 1 MCP at different operating voltages. For each simulation run, we randomly beam 1000 electrons onto input face of MCP and record the output electron's data in a root file. By analyzing the number of electrons in the root file, we can figure out the average gain at each specify operating voltage. We sweep the voltage range from 700 volts to 1200 volts and obtain the voltage-gain curve, as shown in Figure~\ref{fig:G_V_1}. Similarly, we also simulate \# 2 MCP with the same SEY parameters and SEES parameters. The resulting gain-voltage curve for \#2 is shown in Figure~\ref{fig:G_V_2}.

\begin{figure}[htbp]
	\centering
	\includegraphics[width=0.5\textwidth]{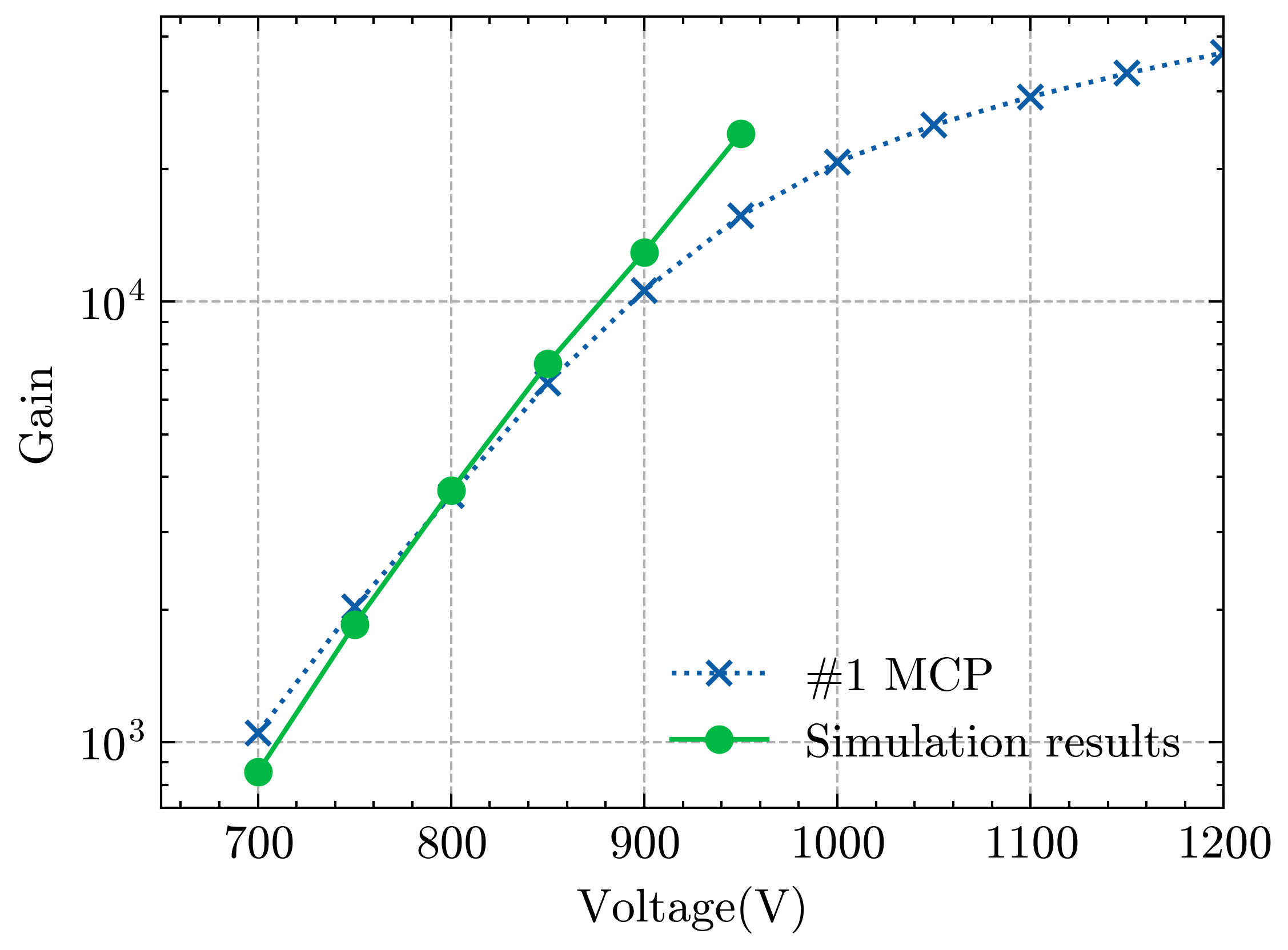}
	\caption{The gain of \#1 MCP versus applied voltage. The blue ``$\times$'' markers denote the experimental measurements and the green points denote the simulated results.}
	\label{fig:G_V_1}
\end{figure}

\begin{figure}[htbp]
	\centering
	\includegraphics[width=0.5\textwidth]{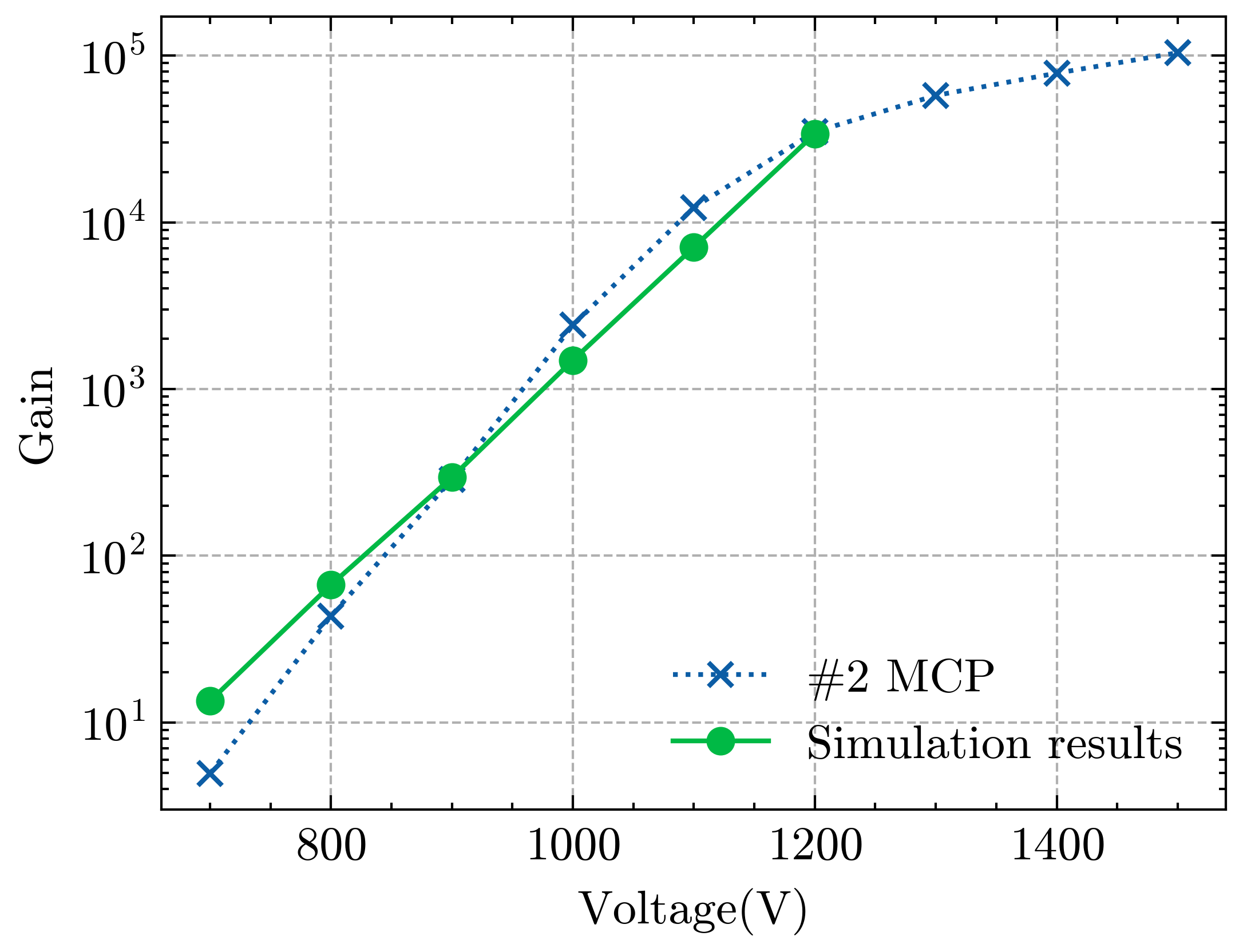}
	\caption{The gain of \#2 MCP versus applied voltage. The blue ``$\times$'' markers denote the experimental measurements and the green points denote the simulated results.}
	\label{fig:G_V_2}
\end{figure}

The two pieces of MCPs are simulated at different applied voltages from 700~V to 1200~V with a step of 100~V using the same SEY and energy spectra of the secondary electrons. For each voltage, the gain is simulated for 1000 times with random primary incident electrons and the average is determined to be the gain at this voltage. The gains with respect to the voltages are shown in Fig.~\ref{fig:G_V_1} and \ref{fig:G_V_2}.

%From the two curves presented above, we can find that there is a significant exponential relationship between gain and applied voltage. It is consistent with other reference paper\cite{adams1966mechanism}. However, this exponential relationship also leads to a computational challenge. As the applied voltage increases, the number of particles that need to be simulated and calculated also increase exponentially. 

%Furthermore, we find that the Monte Carlo results are in good agreement with experimental data. However, when a sufficiently high operating voltage is applied, the MCP is under saturation effect, leading to a suppression of the gain. At this point, the gain and voltage no longer follow an exponential relationship and tend to stabilize at a constant value. In Geant4 Monte Carlo simulation, the space charge effect is not taken into account. As a result, the simulated gain continues to increase as the voltage rises exponentially. From the Figure~\ref{fig:G_V_1} and~\ref{fig:G_V_2}, it is obvious that the gap between experimental data and simulation data become larger in high voltage region.

According to the simulation results shown in Fig.~\ref{fig:G_V_1} and Fig.~\ref{fig:G_V_2}, the gain will increase exponentially with the increasing applied voltage, which is consistent with the experimental measurements and other studies~\cite{evans1965low}, while the measured gain will be obviously lower than the simulated result and tend to stabilize to a constant value when the applied voltages are sufficiently high since the multiplication will be suppressed by the saturation effect~\cite{adams1966mechanism, harris1971saturation} that is not taken into consideration in the simulation at present.

%In addition to the straight channel MCP, we also do some simulations on curved channel MCP. The model is showed in Figure~\ref{fig:curve_MCP}. These MCPs are designed to effectively reduce after-pulses\cite{evans1965low}, which are the main source of noise in high-sensitivity experiments\cite{juno2022juno, marzec2022neutrino, jiang2020study}. In our study, We investigated various geometry parameters, such as channel curvature radius, channel length-to-radius ratio and channel pore radius. Similar to the previous simulations, we beamed 1000 electron onto MCP input face and recorded output electrons data. The resulting gain-voltage curve for the curved channel MCP is shown in Figure~\ref{fig:curve_G_V}. As expected, the gain still increase with operating voltage following an exponential law. 

\begin{figure}[htbp]
	\centering
	\includegraphics[width=0.5\textwidth]{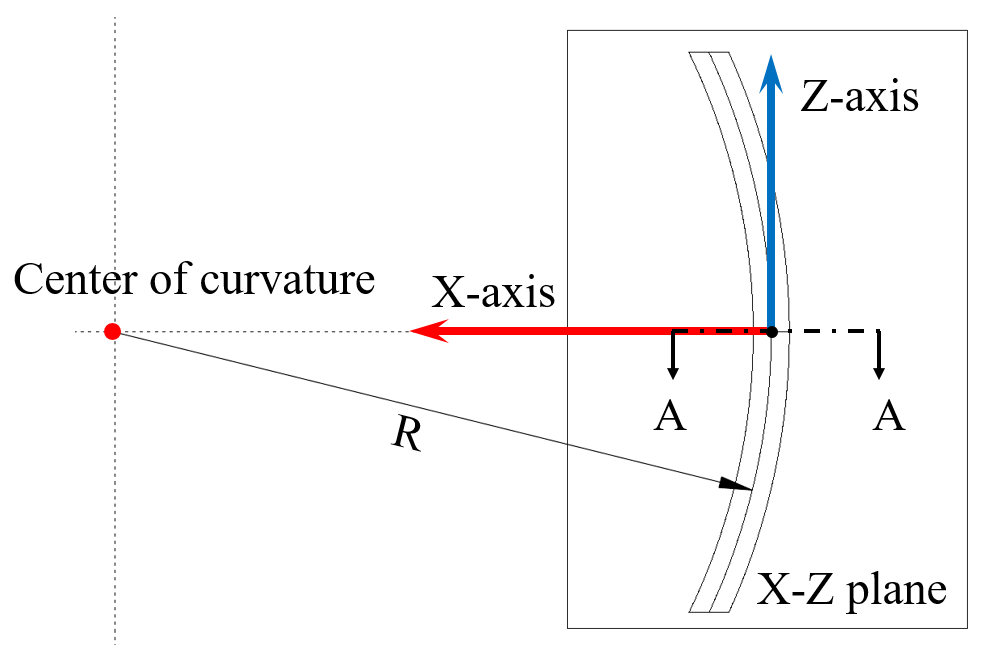}
	\caption{Sectional view of the X-Z plane for a single curved channel. The red point in the left represents the center of the curvature and R denotes the radius of the curvature.}
	\label{fig:topview}
\end{figure}

\begin{figure}[htbp]
	\centering
	\includegraphics[width=0.5\textwidth]{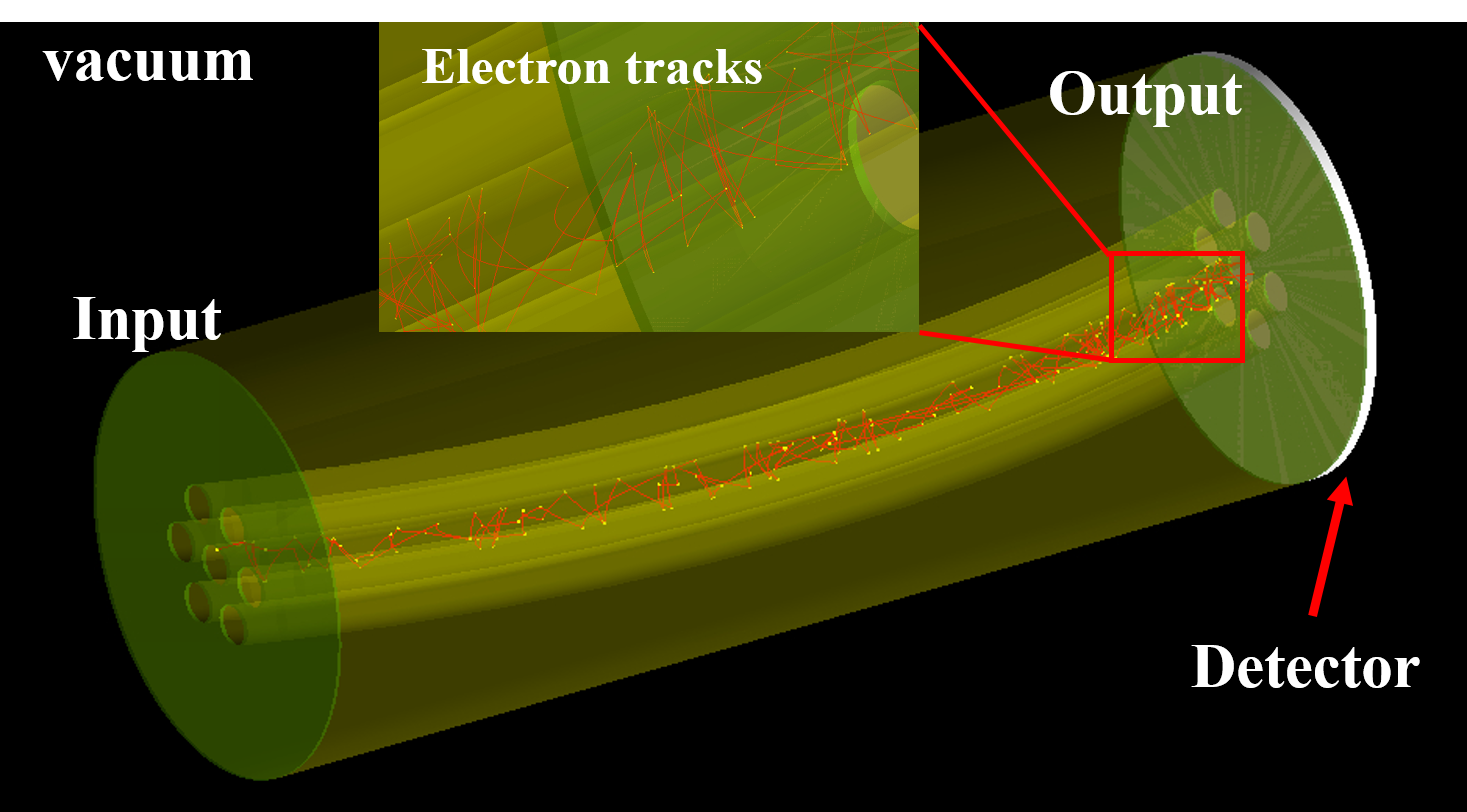}
	\caption{Simulation of the curved-channel MCP. Only part of the electron tracks are displayed.}
	\label{fig:curve_MCP}
\end{figure}

\begin{figure}[htbp]
	\centering
	\includegraphics[width=0.8\textwidth]{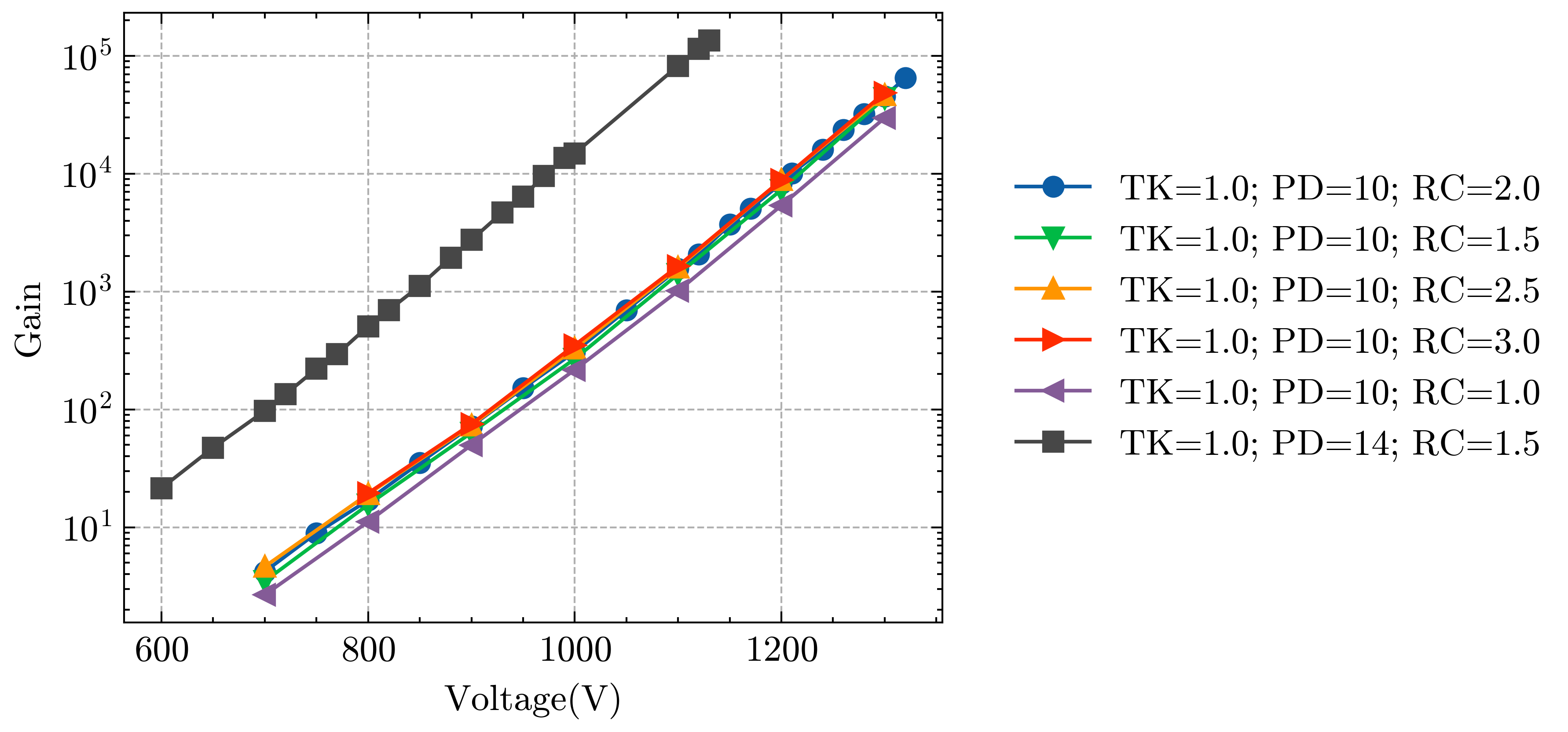}
%	\caption{The relationship of gain and applied voltage under different geometry parameters. TK represents MCP thickness, the unit is millimeter. HR denotes MCP channel pore radius, its unit is micrometer. RT denotes curved channel radius, its unit is millimeter.}
	\caption{Gain with respect to the applied voltage with different geometric parameters. TK represents the thickness of MCP (mm), PD denotes the pore diameter of the channel ($\mu$m), and RC is the radius of curvature of the curved channel (mm).}
	\label{fig:curve_G_V}
\end{figure}

In addition to the straight-channel MCP, the curved-channel MCP, originally designed to effectively reduce the after-pulses~\cite{evans1965low} that is proved to be the main source of noises in the high-sensitivity experiments~\cite{juno2022juno, marzec2022neutrino, jiang2020study}, is also simulated. Fig.~\ref{fig:topview}, viewed from Y-axis, illustrates the bending of the curved channel in a coordinate system. In Geant4, as shown in Fig.~\ref{fig:curve_MCP}, the electron multiplication is simulated in a simplified MCP model with 7 curved channels, of which various geometric parameters are investigated, including the curvature of the curved channel, length-to-diameter ratio and pore diameter of the channels. The same simulation strategy as the straight-channel MCP is applied in multiple conditions. As shown in Fig.~\ref{fig:curve_G_V}, the gain increases exponentially with respect to the applied voltage as expected.

%By varying the curvature radius of channel, we simulated the MCP gain at different operating voltages. The simulation results, as shown in Figure~\ref{fig:curve_G_V}, indicate that the curvature radius of channel has little effect on gain. However, the pore radius has a significant effect on it. To further understand the impact of the pore radius on MCP gain, we simulate the MCP gain with different pore radius while keeping the MCP thickness and operating voltage fixed. Figure~\ref{fig:HR_gain} demonstrates that there exits an optimal pore radius that maximizes MCP gain.

\begin{figure}[htbp]
	\centering
	\includegraphics[width=0.8\textwidth]{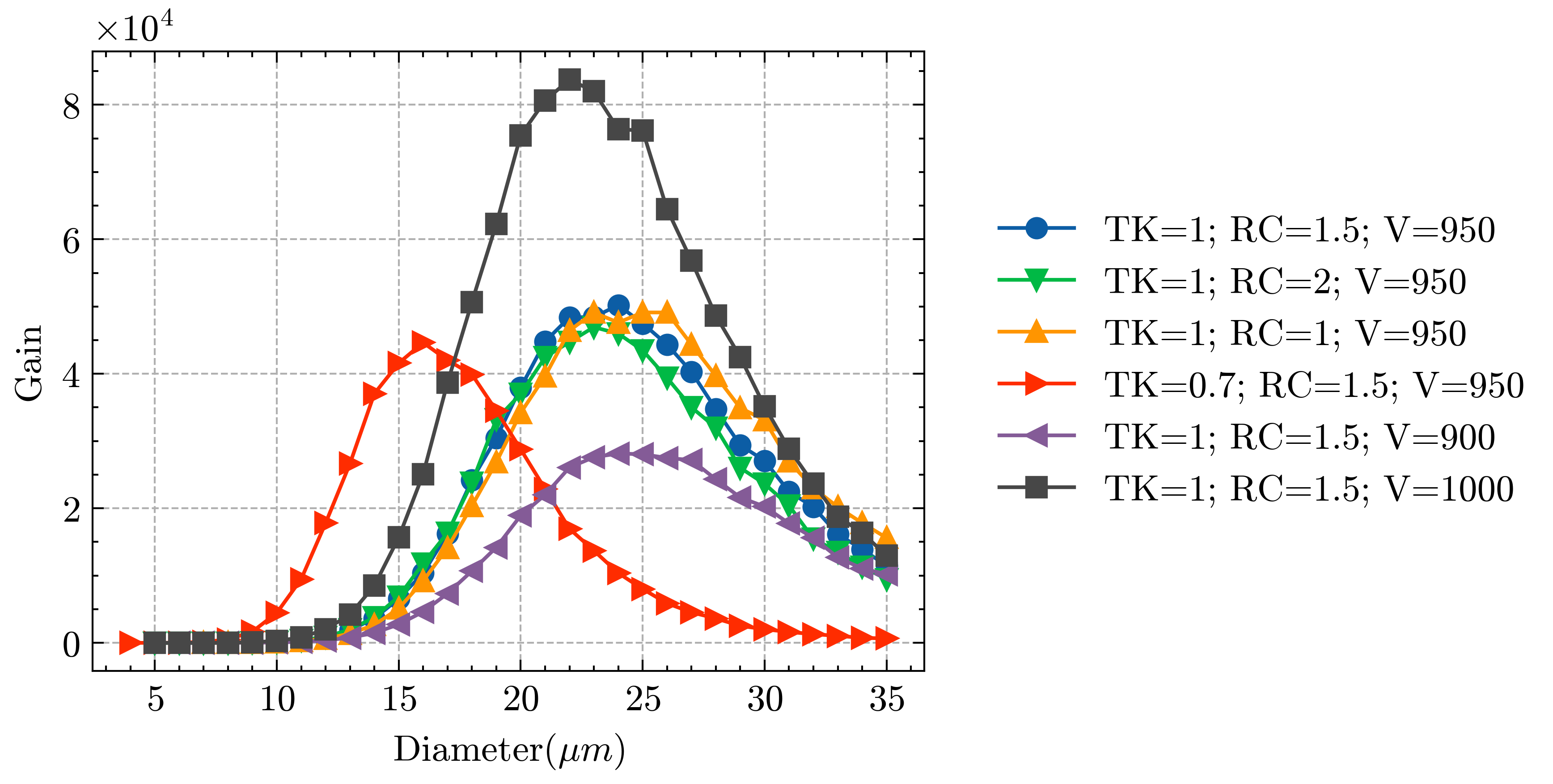}
	\caption{The gain dependence on pore diameter for various geometrical MCP parameters and applied voltages. TK represents the thickness of MCP (mm), and RC is the radius of curvature of the channel (mm). V denotes the applied voltage (V).}
	\label{fig:HR_gain}
\end{figure}

According to the results in Fig.~\ref{fig:curve_G_V}, the curvature of the curved channel has little effect on the gain, but the pore diameter will influence the gain significantly. To further understand the impact of pore diameter, the gain is obtained with different pore diameter when the thickness of MCP and the applied voltage are fixed. As shown in Fig.~\ref{fig:HR_gain}, there is an optimal pore diameter maximizing the gain of MCP.

%We can use a simple model to illustrate the variation trend of gain with pore radius when the MCP thickness and operating voltage are fixed. We can consider the electrons avalanche in the MCP channel and the incident points are continuous distribution along the inner channel. To simplify the problem \cite{adams1966mechanism, eberhardt1979gain}, we regard the MCP multiplication process as discrete dynode multiplication process. We assume that the MCP is equivalent to a PMT with N dynodes, and the average SEY is represented by $\delta$. Therefore, it can approximately calculate the MCP gain as $G = \delta ^ N$. The MCP pore radius can effect both N and $\delta$. When the pore radius is expanded, the electrons will take a longer time to travel from one side of the inner wall to the other side. It leads to electron move further in the axial direction. While, the number of dynodes N will decrease, and the electrons obtain more energy from the electric field, leading to an increase in incident energy, the average $\delta$ value becomes bigger. So an increase in pore diameter affects both N and $\delta$. Figure~\ref{fig:HR_gain} demonstrates that as the pore radius increase, the MCP gain also increase until it reaches the peak. At this point, the growth of $\delta$ dominates above equation reference to $G$, leading to an increase in gain. After MCP gain reaches the peak, as the pore radius continues to increase, the decrease in variable N becomes dominant in above formula about $G$, resulting in a decrease in MCP gain.

The variation trend of the gain in Fig.~\ref{fig:HR_gain} can be illustrated using a simple model. To simplify the issue, the gain of the continuous multiplication process can be equivalently described by the average times of multiplications $N$ and the average SEY $\delta$~\cite{adams1966mechanism, eberhardt1979gain}, so that the gain can be approximately calculated as $G = \delta ^ N$. The pore diameter of MCP affects both $N$ and $\delta$. When the pore diameter becomes larger, the produced secondary electrons will fly for a longer distance along the direction of the electric field before hitting the inner wall of the channel and starting the next multiplication, which will decrease $N$ due to the relatively less times of multiplications before flying out of the channel and increase $\delta$ due to the more energy the electrons can get. When the pore diameter continuously increases, the increasing of $\delta$ dominates the multiplication before reaching the maximum so that the gain rises, and then the decreasing of $N$ dominates the multiplication, making the gain fall down after the maximum.

%In this section, we use Geant4 to simulate both straight channel MCPs and curved channel MCP. The straight channel MCP simulation mainly aimed to calibrate SEEC parameters. Through multiple iterations, we obtained appropriate parameters that could accurately describe the secondary electron emission properties of the MCP's inner wall material. For curved channel MCP, we studied the influence of geometric parameters on the gain. In conclusion, the curvature radius has little effect on the MCP gain, allowing us to increase the channel curvature to suppression after-pulses. Additionally, the research of pore radius shows that when the MCP thickness and operating voltage are fixed, adjusting the pore radius can effectively optimize the MCP gain. 
 
In summary, the curvature of the curved channel affects little on the gain of an MCP, allowing us to increase the curvature to suppress the after-pulses. Our work also proves that there is an optimal pore diameter that maximizes the gain when the thickness and applied voltage of an MCP are fixed.

\subsection{Electron tracks inside the MCP channel}
\label{sec:track_simulation}

%In fact, as a Monte Carlo simulation software framework, Geant4's main task is to track the trajectories of particles and sample various types of reactions. This allows us to easily to analyze electrons behavior in the MCP channel, which is nearly impossible to observe in experiment. We set parameters for the curved channel MCP as listed in Table~\ref{tab:curve_mcp_1_par}, and apply a voltage of 950V. When a single electron incident in the MCP channel, we can obtain over $10^5$ track data by the action hook class in Geant4. According to these track data, we can uncover some novel insights. 

The multiplication process is simulated track by track in this work, allowing us to analyze the behavior of the secondary electrons inside the channel, which is nearly an impossible task in experiments. The geometric parameters of the simulated curved-channel MCP are listed in Table~\ref{tab:curve_mcp_1_par} and the applied voltage is 950~V. Over $1\times10^5$ secondary electron trajectories are recorded for a single incident, which shed light to uncover some novel insights.

\begin{table}[htbp]
	\centering
	\doublespacing
	\caption{The geometric parameters and applied voltage of the simulated curved-channel MCP.}
	\begin{tabular}{cc}
		\hline
		\hline
		Parameter/Unit   &   Curved channel MCP  \\
		\hline
		\texttt{MCP thickness/mm} &   1 \\
		\texttt{Pore diameter/$\upmu$m}    &   20 \\
		\texttt{Applied voltage/V}  &   950 \\
		\texttt{Radius of curvature of the channel/mm}  &   1 \\
		\hline
		\hline
	\end{tabular}
	\label{tab:curve_mcp_1_par}
\end{table}

%From Figure~\ref{subfig:z_direction}, we can find that even under the 950V operating voltage, the average axial jump distance of electrons is approximately 20 $\upmu \mathrm{m}$. It's a  deviation from what was previously  believed. Many papers\cite{eberhardt1979gain, harris1971saturation} compare the MCP multiplication process to a dynode PMT, where most of electrons  undergo about 10 times multiplication steps, with each step SEY value is about 2-3. However, our simulation results show that the electron jump distance in channel is about 20 $\upmu \mathrm{m}$. In a 1 mm thick MCP, the total number of multiplications is about 50. To achieve a gain of $10^5$, the average electron  yield is about 1.1.

\begin{figure}[htbp]
	\centering
	\subfigure[Jump distance of the secondary electrons along the axis of the channel.]{
		\includegraphics[width=0.4\textwidth]{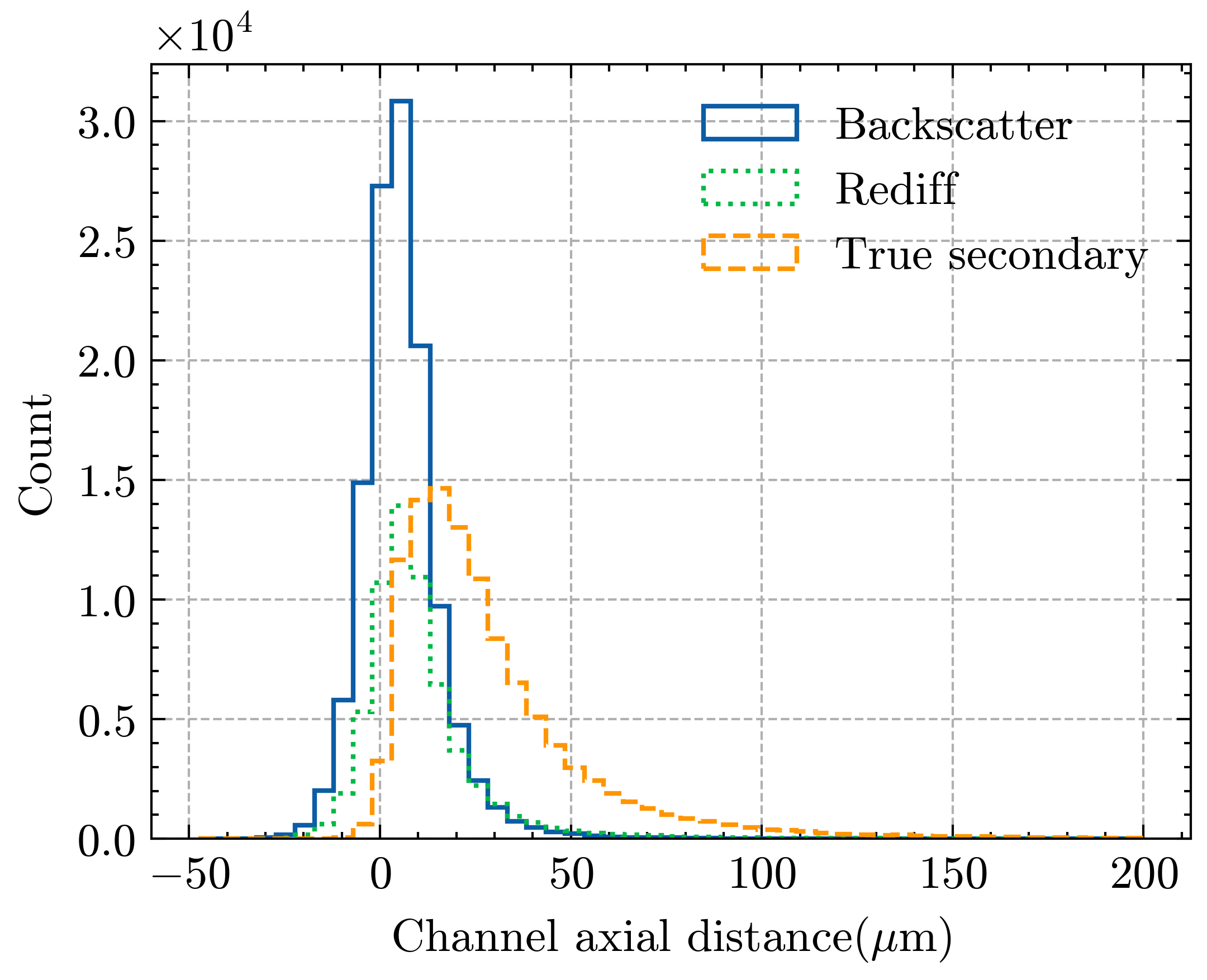} 
		\label{subfig:z_direction}
	} 
	\subfigure[Energy of the electrons hitting the inner wall of the channel.]{
		\includegraphics[width=0.4\textwidth]{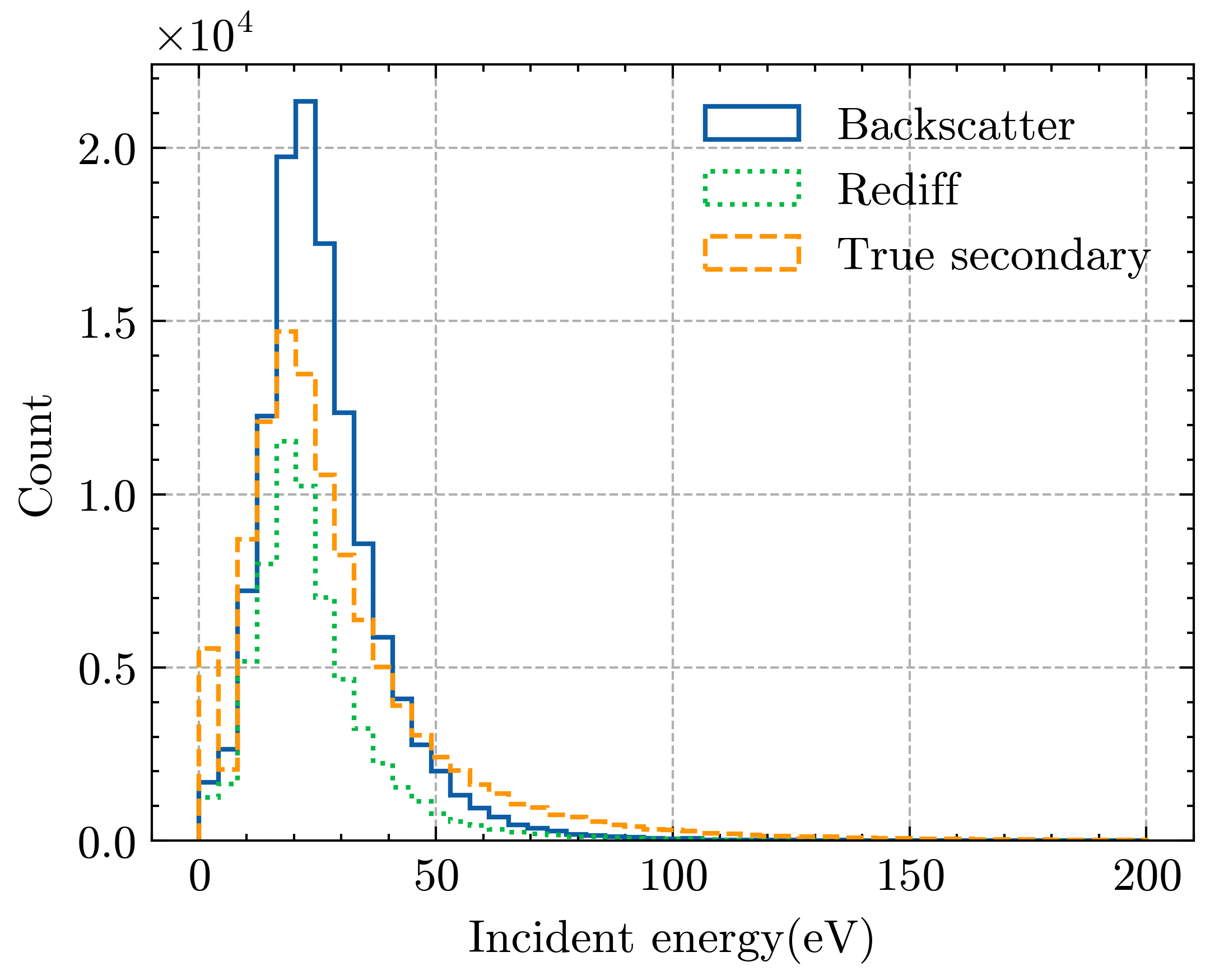}
		\label{subfig:incident_energy}
	}
	\subfigure[Initial energy of the secondary electrons.]{
		\includegraphics[width=0.4\textwidth]{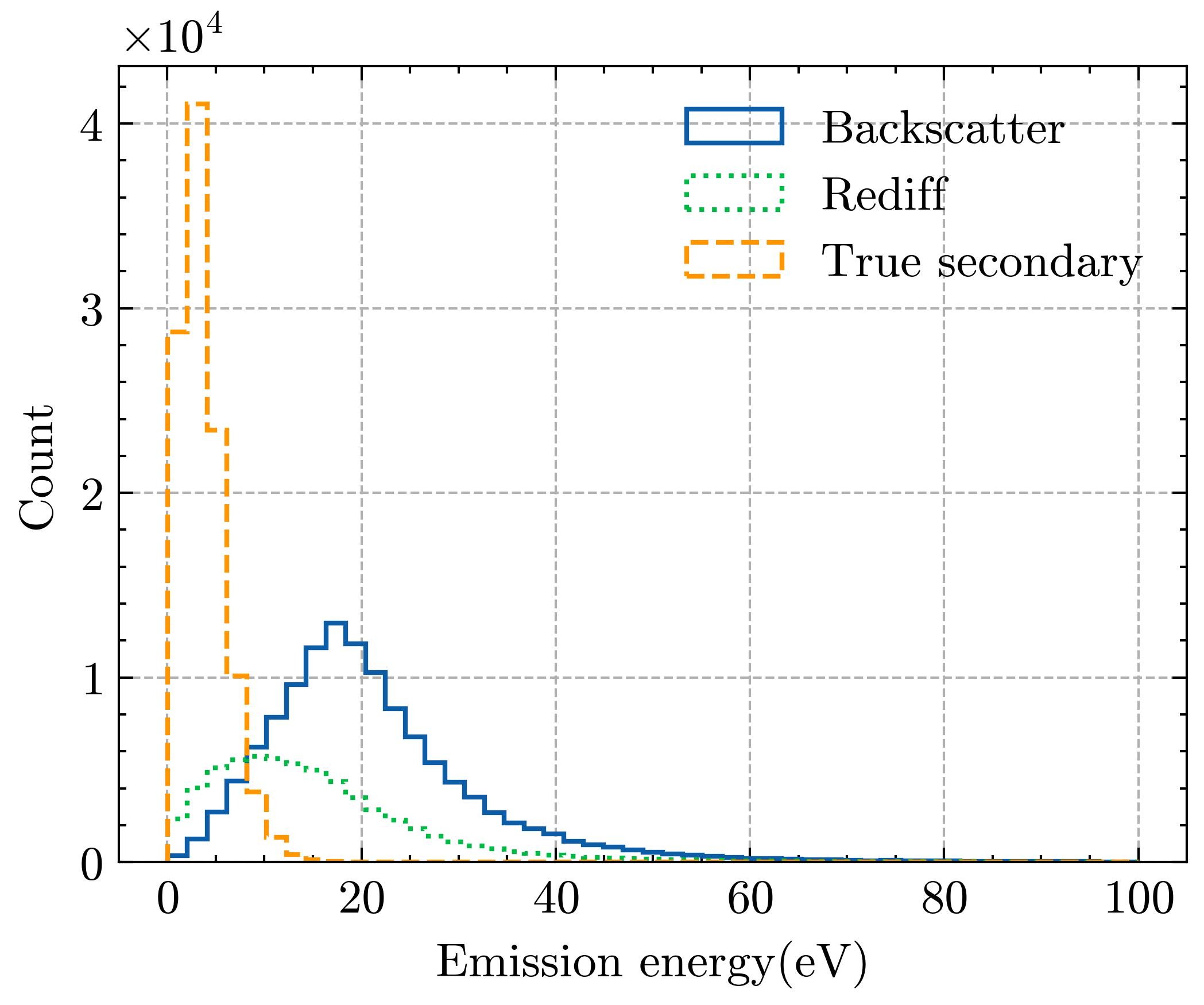}
		\label{subfig:vertex}
	}
	\subfigure[Direction of the secondary electrons.]{
		\includegraphics[width=0.4\textwidth]{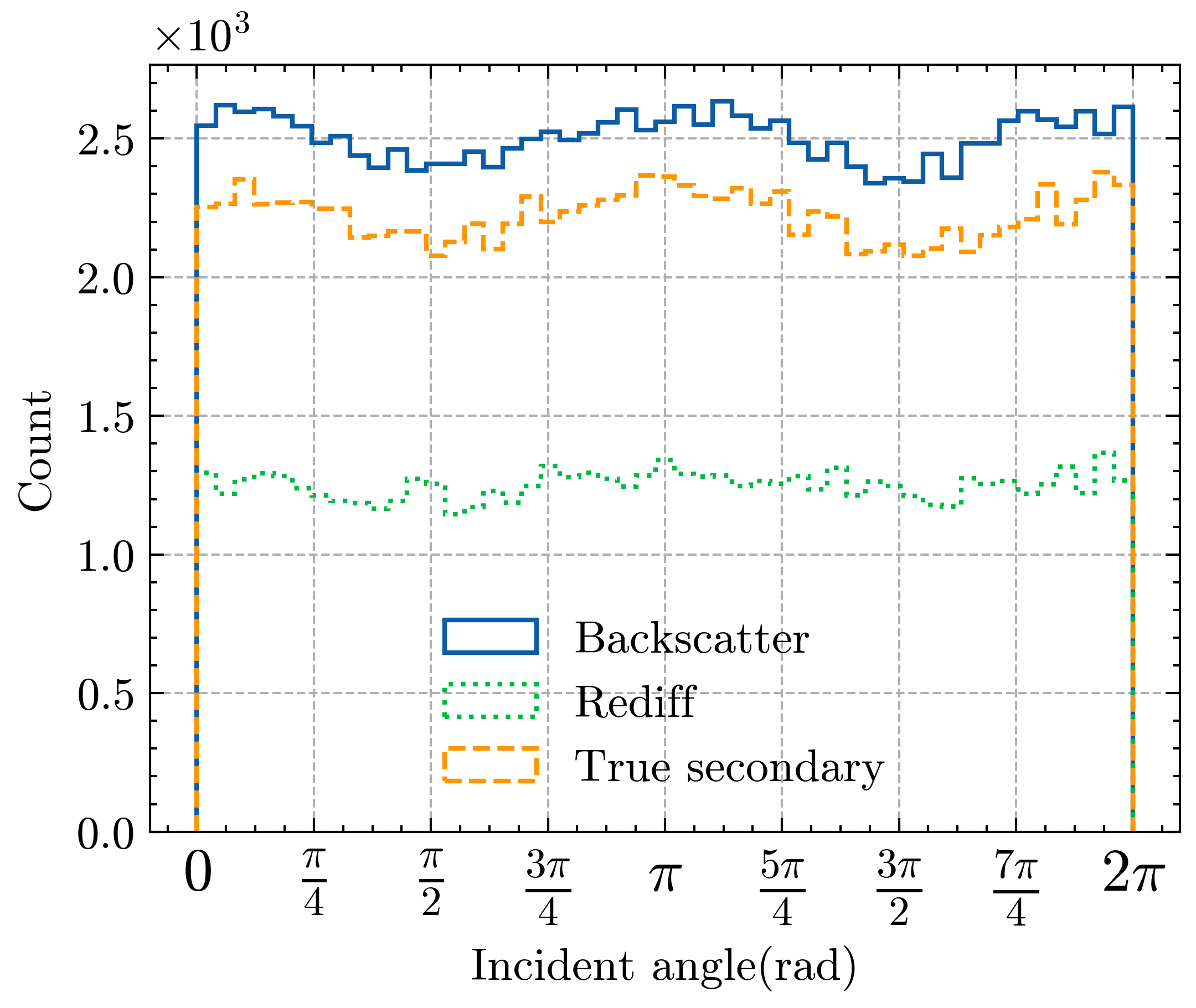}
		\label{subfig:angle}
	}
	
	\caption{Typical distributions of the secondary electrons inside the channel.}
	\label{fig:track}
\end{figure}

From Fig.~\ref{subfig:z_direction}, the average jump distance of the secondary electrons along the axis of the channel is approximately 20~$\mu$m, so that the electrons multiply for about 50 times for an MCP of 1~mm thickness, and the average SEY should be around 1.1 to achieve a gain of $10^5$. The result is different from what was previously said in many papers~\cite{eberhardt1979gain, harris1971saturation} that the electrons multiply for about 10 times with the average SEY about $2\sim3$.

%Figure~\ref{subfig:incident_energy} shows the incident energy of electrons in channel, which is as excepted to be around 30 eV on average. This energy is slightly higher than the threshold energy $\delta = 1$  in SEY curve. According to the Figure~\ref{fig:modified_SEY}, these electros are  most likely  to take participate in backscattering. Only a small proportion electrons have a significant effect on the multiplication process.

Fig.~\ref{subfig:incident_energy} shows that the incident energy of the electrons is about 30~eV on average, slightly higher than the threshold energy when $\delta=1$ in the SEY curve. According to Fig.~\ref{fig:modified_SEY}, these electrons are most likely to participate in backscattering and only a small proportion of the electrons have a significant influence on the multiplication.

%The vertex distribution in Figure~\ref{subfig:vertex} shows the energy of electrons emitted from the MCP inner wall. There are three peaks. The left side peak corresponds to the true secondary electrons, as the true secondary electron emitted energy is approximately independent of the incident energy. The right side wider peak corresponds to the contribution from backscatter electrons. The peak that is sandwiched between above two peaks corresponds to rediffused electron.

Fig.~\ref{subfig:vertex} shows the initial energy of the secondary electrons. The narrow yellow peak corresponds to the true secondary electrons that are approximately independent of the incident energy. The blue, wider peak is the contribution from the backscattered electrons. The green curve represents the rediffused electrons.

%Lastly Figure~\ref{subfig:angle} suggests that the electron motion in the channel is not isotropy. We can find that electrons tend to bombard direction parallel to the plane where channel bends. This tendency leads to the output electron distribution in a elliptic, rather than a circular distribution.
\begin{figure}[htbp]
	\centering
	\includegraphics[width=0.7\textwidth]{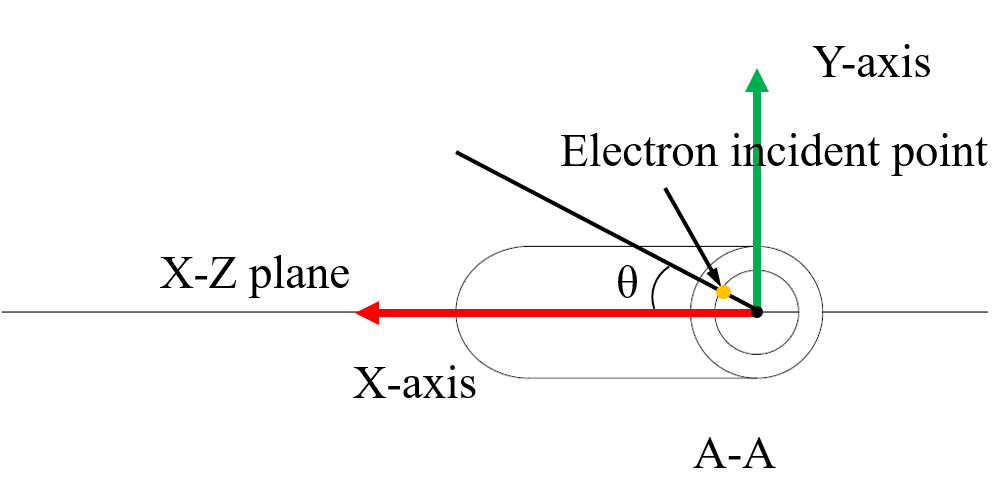}
	%%\caption{The diagram of angele parameters $\theta$, which is described in ~\ref{subfig:angle}}
	\caption{The sectional view of the X-Y plane for a single curved channel. }
	\label{angle_diagram}
\end{figure}

To study the spatial orientation of the electrons inside a channel during the multiplication, an angle $\theta$ is defined as shown in Fig.~\ref{angle_diagram}. The X-direction corresponds to $\theta = 0$ and the Y-direction corresponds to $\theta = \pi/2$. Fig.~\ref{subfig:angle} shows the distribution of the impact positions of the electrons when they hit the inner wall of the channel, which are described by the angle $\theta$. It suggests that the electron motion inside the channel is not isotropic. They tend to concentrate a little towards the X-Z plane.

%In this section, we research the electrons behavior in MCP channel by Geant4. By analyzing the simulation results, it shows that the electrons average axial jump distance is shorter than what we excepted before. At the same time, to achieve the gain that is measured in experiment. Most of electrons undergo 50-60 times collision for a MCP whose length-diameter ratio is around 50. To sum up, the electrons in MCP are in a state, which is the majority electrons are not accelerate fully, and the average yield that these electron is only a slightly larger than 1.

\section{Conclusion}
\label{sec:conclusion}

%According to our simulation result, we can find that the curved channel MCP gain increase with applied voltage exponentially before reaching to saturation. Additionally, when the MCP thickness and operating voltage are fixed, the MCP gain is significantly affected by the diameter of MCP pores . For instance, when MCP applied 950V and the thickness of MCP is 1 mm. It's optimum diameter is about 20 $\upmu \mathrm{m}$.

According to our simulation, the gain of the curved-channel MCP increases exponentially with the increasing applied voltage before reaching saturation. Additionally, the gain is significantly affected by the pore diameter of the channel when the thickness and applied voltage are fixed. When the applied voltage and the thickness of an MCP are 950~V and 1~mm, correspondingly, the optimum pore diameter is about 20~$\mu$m. The curvature of the curved channel has no significant effect on MCP gain.

%Furthermore, by analyzing the electron track data in the MCP channel (as shown in  Figure~\ref{fig:track}), we can conclude that the acceleration distance of electrons in the channel is so short that leads to particle energy that bombard channel inner wall is low. Consequently, only a few electrons can generate true secondary electrons, while the majority of them undergo backscatter. From this perspective, enlarging the pore diameter can increase the MCP gain. In fact, under fixed thickness and operating voltage conditions, we can determine the optimal pore diameter by using the simulations. 

Furthermore, we found that the accelaration distance of the secondary electrons inside the channel is too short to obtain enough energy before bombarding the inner wall of the channel, so that only a few electrons are able to produce the true secondary electrons and most of them are backscattered. Based on this perspective, enlarging the pore diameter within a reasonable range can increase the gain and we can also determine the optimal pore diameter of an MCP with fixed thickness and applied voltage by simulation.

There are still issues needed to be considered in the further simulation studies, including the after-pulse of MCP, saturation effect and so on. Additionally, a comparison between chevron MCP and curved-channel MCP could be conducted. To achieve these goals, we have to resolve the challenges about the time consumption of calculation, and the design of the algorithm and simulation models, which will be the concentration of our next-stage work.
%Further study is still required for the simulation, including the after-pulse of MCP, saturation and some other signification characters. Additionally, a comparison between chevron MCP and curved-channel MCP can be conducted. To deal with above problems, we have to resolve the challenge about calculation time, algorithm and model design. We will concentrate on these challenge and do some try in the future.

\section*{Acknowledgements}
This work is supported by the National Natural Science Foundation of China (Grant 11975017), the State Key Laboratory of Particle Detection and Electronics (SKLPDE-ZZ-202215). We are gratitude to Dr. Kaile Wen for his insightful discussions.



\bibliographystyle{elsarticle-num}
\bibliography{references}


%% \begin{thebibliography}{00}


%% \bibitem{}

%% \end{thebibliography}
\end{doublespacing}
\end{document}